\documentclass[preprint]{emulateapj-rtx4}

\shorttitle{Host galaxies of SDSS-RM quasars}
\shortauthors{Matsuoka et al.}

\begin{document}


\title{The Sloan Digital Sky Survey Reverberation Mapping Project: \\Post-Starburst Signatures in Quasar Host Galaxies at {\scriptsize z} $< 1$}


\author{Yoshiki Matsuoka\altaffilmark{1,2}, 
	Michael A. Strauss\altaffilmark{2}, 
	Yue Shen\altaffilmark{3,4,$\dagger$},
         William N. Brandt\altaffilmark{5,6,7}, 
	Jenny E. Greene\altaffilmark{2},\\
	Luis C. Ho\altaffilmark{3,8},
         Donald P. Schneider\altaffilmark{5,6},
	Mouyuan Sun\altaffilmark{5,6,9},
	Jonathan R. Trump\altaffilmark{5,6,$\dagger$} 
	}


\altaffiltext{1}{National Astronomical Observatory of Japan, Mitaka, Tokyo 181-8588, Japan; yk.matsuoka@nao.ac.jp}
\altaffiltext{2}{Princeton University Observatory, Peyton Hall, Princeton, NJ 08544, USA}
\altaffiltext{3}{Kavli Institute for Astronomy and Astrophysics, Peking University, Beijing 100871, China}
\altaffiltext{4}{Carnegie Observatories, 813 Santa Barbara Street, Pasadena, CA 91101, USA} 
\altaffiltext{5}{Department of Astronomy \& Astrophysics, The Pennsylvania State University, University Park, PA 16802, USA}
\altaffiltext{6}{Institute for Gravitation \& the Cosmos, The Pennsylvania State University, University Park, PA 16802, USA}
\altaffiltext{7}{Department of Physics, The Pennsylvania State University, University Park, PA 16802, USA}
\altaffiltext{8}{Department of Astronomy, School of Physics, Peking University, Beijing 100871, China}
\altaffiltext{9}{Department of Astronomy and Institute of Theoretical Physics and Astrophysics, Xiamen University, Xiamen, Fujian 361005, China}
\altaffiltext{$\dagger$}{Hubble Fellow}


\begin{abstract}

Quasar host galaxies are key for understanding the relation between galaxies and the supermassive black holes (SMBHs) at their centers.
We present a study of 191 broad-line quasars and their host galaxies at $z < 1$, using high signal-to-noise ratio (SNR) spectra
produced by the Sloan Digital Sky Survey Reverberation Mapping project.
Clear detection of stellar absorption lines allows a reliable decomposition of the observed spectra into nuclear and host components,
using spectral models of quasar and stellar radiations as well as emission lines from the interstellar medium.
We estimate age, mass $M_*$, and velocity dispersion $\sigma_*$ 
of the host stars, the star formation rate (SFR), quasar luminosity, and SMBH mass $M_\bullet$, for each object.
The quasars are preferentially hosted by massive galaxies with $M_* \sim 10^{11} M_\odot$ characterized by stellar ages around a billion years, 
which coincides with the transition phase of normal galaxies from the blue cloud to the red sequence.
The host galaxies have relatively low SFRs and fall below the main sequence of star-forming galaxies at similar redshifts.
These facts suggest that the hosts have experienced an episode of major star formation sometime in the past billion years, which was
subsequently quenched or suppressed.
The derived $M_\bullet - \sigma_*$ and $M_\bullet - M_*$ relations agree with our past measurements and are consistent with no evolution from the
local Universe.
The present analysis demonstrates that reliable measurements of stellar properties of quasar host galaxies are possible with high-SNR fiber spectra, 
which will be acquired in large numbers with future powerful instruments such as the Subaru Prime Focus Spectrograph.

\end{abstract}




\section{Introduction \label{sec:intro}}

Over the last few decades, it has become clear that 
supermassive black holes (SMBHs) are ubiquitous at galaxy centers throughout the Universe.
Stellar or gaseous motions at galactic centers indicate that almost every galaxy bulge in the local Universe harbors a SMBH
\citep[e.g., ][]{dressler89,kormendy93,magorrian98}.
The correlation between SMBH mass ($M_\bullet$) and the stellar velocity dispersion ($\sigma_*$) or mass of the bulges 
\citep[e.g., ][]{merritt01,mclure02,haring04,gultekin09} suggests that the two coevolve, or at least strongly influence each other.
The observed scatter in the $M_\bullet - \sigma_*$ relation is remarkably small
\citep{ferrarese00,gebhardt00}.
This situation may require a fine-tuning process, which is perhaps a combination of internal physics within galaxies or dark halos
and mass averaging via frequent mergers 
\citep[see the recent review by][and references therein]{kormendy13}.
Beyond the local Universe, SMBHs are known to exist even at very high redshifts.
Recent wide-field surveys 
have discovered more than 40 quasars at $z > 6$ \citep[e.g.,][]{fan06,jiang09, willott10,banados14}, with the most distant discovery at $z = 7.1$ \citep{mortlock11}.
The estimated SMBH mass exceeds $M_\bullet = 10^9 M_\odot$ in many cases \citep[e.g.,][]{jiang07, wu15},
which has initiated intense discussion on the process of their formation and evolution in the short time scales of the early Universe \citep[e.g.,][]{ferrara14,madau14}.
The $M_\bullet - \sigma_*$ relation of these objects tend to be offset toward higher $M_\bullet$ compared to the local relation at fixed $\sigma_*$ \citep[e.g.,][]{wang10}, 
but possible biases in sample selection and measurements complicate the interpretation \citep[see, e.g.,][]{schulze11, shen13}.

Since the \citet{soltan82} argument and the latest observations suggest that the local SMBHs have acquired most of their mass during 
an active galactic nucleus (AGN) phase \citep[e.g.,][]{yu02, marconi04, brandt15, comastri15}, 
a key to understanding SMBH growth and the origin of the $M_\bullet - \sigma_*$ relation may be
found in observational properties of AGNs.
One of the outstanding issues is the nature of their host galaxies;
in which galaxies do AGNs occur, and what impact do they have on the galaxies? 
These questions have been addressed by numerous studies in recent years, especially in terms of negative feedback effects of AGNs on the star formation rates (SFRs) 
of the host galaxies.
It has been suggested that the energy and/or momentum input of AGNs, which may be triggered along with starbursts by interactions/mergers of gas-rich galaxies 
\citep[e.g.,][]{barnes91}, could suppress star formation by expelling the cold gas from galaxies 
\citep[e.g.,][]{dimatteo05,springel05} and/or heating the gas in dark halos \citep[e.g.,][]{mcnamara07}.
Indeed, recent observations of AGN-driven gas outflows on small scales \citep[e.g.,][]{pounds03, trump06, ganguly07, reeves09, tombesi10, gofford13}
as well as large scales \citep[e.g., ][]{nesvadba06,feruglio10,greene11,cano-diaz12,liu13a,liu13b} and
extended emission-line regions photoionized by AGNs \citep[e.g.,][]{fu09,husemann10,matsuoka12,keel12} 
strongly suggest that AGNs have a significant impact on their host galaxies.
The idea of negative AGN feedback is also favored because it may provide a solution to longstanding problems with galaxy formation models; 
for example, it is currently the most compelling process to reconcile the quite different shapes of the dark halo mass function predicted 
in the $\Lambda$CDM cosmology and the observed galaxy stellar mass function at the high mass end \citep[e.g.,][]{bower06,croton06,somerville08, dave11,choi11}.
The AGN feedback may have two distinct modes, i.e., quasar mode and radio mode, which become efficient at different phases of SMBH growth
and galaxy evolution \citep[see, e.g.,][]{fabian12}.

There have been a number of studies on host galaxies of various AGN classes. 
Different types of AGNs may play different roles in galaxy evolution \citep[e.g.,][]{hickox09, dipompeo14, dipompeo15}, hence it is crucial to understand 
the entire AGN and host populations.
\citet{elitzur12} pointed out that 
optically-classified narrow-line AGNs are more likely found in dusty circumnuclear environments than are broad-line AGNs \citep[see also][]{netzer15}.
In a merger-driven scenario of galaxy and SMBH co-evolution \citep{hopkins06}, a major merger first creates the dusty starburst/AGN phase observed as 
luminous infrared galaxies, and then evolves into the dust-free phase observed as optically-luminous quasars \citep[see also][]{sanders88}.
Therefore, AGNs selected at different wavelengths may have fundamentally different host properties; for instance, dusty infrared-selected AGNs 
\citep[e.g., ][]{lacy13} may be associated with more active star formation on average than are optically-selected AGNs.

Early studies on this subject \citep[e.g., ][]{boroson82, kotilainen94, ronnback96, bahcall97, mclure99, kirhakos99, dunlop03, jahnke04, jahnke07} 
have shown that 
host galaxies of optically-luminous AGNs
are not a random subsample of normal galaxies. 
Instead, these AGNs were found to reside in luminous galaxies with young stars and a higher than usual fraction of close companions or tidal features.
It has also been suggested that the host galaxies have a substantial contribution from intermediate-age stars \citep[e.g., ][]{canalizo13},
which may correspond to the late stage of accretion episode predicted in the above merger-driven evolutionary scenario.
Recent studies have taken advantage of a large sample of obscured AGNs,
which show no visible sign of accretion-disk emission or broad lines at ultraviolet (UV) to optical wavelengths,
 discovered in large spectroscopic surveys
\citep[e.g.,][]{kauffmann03, zakamska03, hao05, reyes08} or X-ray observations \citep[e.g.,][]{nandra07, georgakakis08, silverman08, mullaney12}.
Dust extinction serves as a natural coronagraph in these objects and makes analysis of the host galaxies 
much easier than in unobscured populations.
Overall, these studies have statistically confirmed the earlier findings and shown that obscured AGNs reside preferentially in massive galaxies.
While there have been controversies over the host colors, recent work points out the importance of using a mass-matched control sample of 
inactive galaxies for unbiased measurements. 
Once the mass dependence is properly accounted for, the AGN fraction is more or less constant across the host colors or moderately enhanced in bluer galaxies
\citep[e.g.,][]{silverman09, xue10, hainline12, rosario13, trump15}.
At the same time, it has been demonstrated that the color by itself may not be a good tracer of the star-forming properties of the host galaxies 
\citep[e.g.,][]{cardamone10, rosario13}.

As a complementary effort to the studies of obscured AGNs in large contemporary surveys, we are investigating the
host galaxies of ``classical" broad-line quasars selected at optical wavelengths.
In our previous work \citep{matsuoka14}, we analyzed the images of about 1,000 Sloan Digital Sky Survey \citep[SDSS;][]{york00, eisenstein11} 
broad-line quasars at $z < 0.6$ and explored their host properties.
Deep co-added SDSS images of the Stripe 82 region \citep{annis14}  in five bands \citep[$u$, $g$, $r$, $i$, and $z$;][]{fukugita96} were used for the analysis.
We developed a technique to decompose spatially a quasar image into nuclear and host components, using the point spread function (PSF)
and \citet{sersic68} models, and measured the colors and stellar masses of the host galaxies.
We showed that the quasars are hosted exclusively by massive galaxies 
and that these host galaxies are considerably bluer than the red sequence, 
showing stark contrast to the color-magnitude diagram (CMD) of inactive galaxies \citep[see also][]{trump13}.
We also found a positive correlation between $M_\bullet$ and stellar mass $M_*$, but the relation is offset toward large $M_\bullet$ compared to 
the local relation at fixed $M_*$. 
While this result could indicate that SMBHs grow earlier than do the host galaxies, we deferred a definitive conclusion until observational biases are fully understood
\citep[e.g.,][]{schulze11, shen13}.

In this paper, we present the results of a spectral decomposition analysis of SDSS broad-line quasars at $z < 1$.
By making use of the high signal-to-noise ratio (SNR) spectra produced by the SDSS Reverberation Mapping (RM) project (see the following section), we measure 
the spectral properties of about 200 quasar host galaxies and explore their nature.
This paper is organized as follows. 
The data and sample are presented in \S 2. 
\S 3 describes the technique used to decompose a quasar spectrum into nuclear and host components, and the associated error assessments.
The main results appear in \S 4 and are discussed in context of galaxy and SMBH evolution in \S 5.
The summary follows in \S 6.
The cosmological parameters $H_0 = 70$ km s$^{-1}$ Mpc$^{-1}$, $\Omega_{\rm M} = 0.3$, and $\Omega_{\rm \Lambda} = 0.7$ are used throughout this work.
All magnitudes are presented on the AB system \citep{oke83}.

\section{Data acquisition and basic processing \label{sec:data}}

We use the data acquired in the SDSS-RM project, in which a single spectroscopic field was repeatedly observed to explore the variability of quasars.
The full technical details of the project are found in \citet{shen14}, and hence we repeat only the most relevant aspects here.
SDSS-RM was conducted during the dark/grey time in the final season (2013--2014) of the SDSS-III Baryon Oscillation Spectroscopic Survey 
\citep[BOSS;][]{dawson13} following its early completion due to unexpectedly good weather.
The target field (centered at $\alpha_{\rm J2000}$ = $14^{\rm h}14^{\rm m}49^{\rm s}.00$, $\delta_{\rm J2000}$ = $+53^{\circ}05^{\rm m}00^{\rm s}.0$) 
coincides with the Panoramic Survey Telescope \& 
Rapid Response System 1
\citep[Pan-STARRS 1;][]{kaiser10} Medium Deep Field MD07 which lies within the Canada-France-Hawaii Telescope Legacy Survey W3 field.
About 1,000 spectroscopically confirmed quasars are known in this 7-deg$^2$ field from SDSS-I/II and SDSS-III, most of which were targeted by the
BOSS spectroscopy.
Additional quasars have been identified by the DEEP2 Galaxy Redshift Survey \citep{newman13} and by follow-up spectroscopy of Pan-STARRS 1 
variable sources (P. J. Green et al., in preparation).
The redshifts of these quasars have been determined by matching identified emission lines with a pre-defined line list or by cross correlating 
observed spectra with spectral templates.

\begin{figure}
\epsscale{1.1}
\plotone{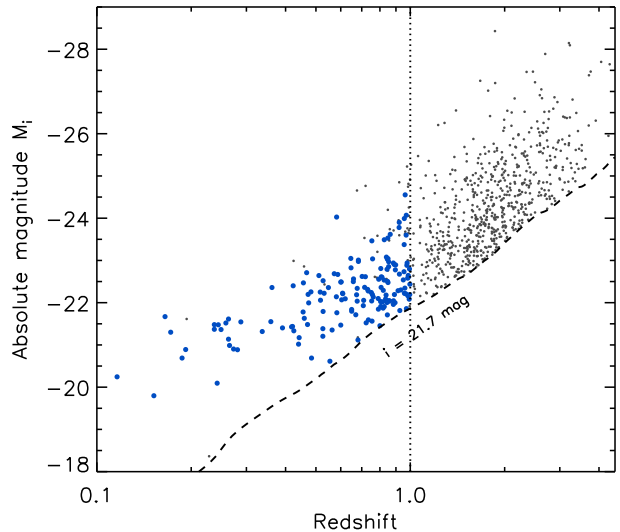}
\caption{Redshifts and $i$-band absolute magnitudes of the 849 SDSS-RM quasars (small and large dots), defined by the flux limit $i < 21.7$ mag (dashed line).
We analyze 191 quasars at $z < 1.0$ (dotted line) in this study.
Our final sample consists of 156 quasars (large dots) with successful spectral decomposition (see \S \ref{sec:application}).
\label{fig:zMi}
}
\end{figure}

The SDSS-RM quasars were selected as a flux-limited sample of 849 objects at $i < 21.7$ mag, 
with objects in fiber collisions being removed.
Each quasar is labelled with an identification number \texttt{RMID}. 
We stress that the entire sample is composed of spectroscopically-confirmed broad-line quasars, the vast majority of which were selected at optical wavelengths;
further details of the sample selection are given in \S 3.1 of \citet{shen14}.
The quasars follow the luminosity-redshift distribution expected from the luminosity function estimates 
in \citet{hopkins07}, suggesting that the sample is fairly uniformly selected above 
the adopted flux limit. 
Figure \ref{fig:zMi} displays the distribution of their redshifts and absolute magnitudes in SDSS $i$ band ($M_i$).
$M_i$ was derived from the SDSS PSF magnitudes, $k$-corrected to $z = 0$ with the correction factors taken from \citet{richards06}.
The quasars analyzed in this work ($z < 1$; see \S \ref{sec:application}) have luminosities comparable to or lower than the classical quasar threshold 
$M_B = -23$ mag \citep{schmidt83}, which corresponds to $M_i \simeq -23$ mag assuming the SDSS quasar composite spectrum of \citet{vandenberk01}
and correcting for the different cosmology used in \citet{schmidt83}.

The SDSS-RM observations were carried out with the BOSS spectrograph \citep{smee13} mounted on the SDSS 2.5-m telescope \citep{gunn06} 
in 32 epochs from January to July 2014.
The spectrograph provides wavelength coverage from $\lambda$ =  3,650 to 10,400 \AA\ with a spectral resolution of $R \sim 2,000$.
The BOSS plates allow 1,000 fibers of 2\arcsec\ diameter to be observed simultaneously;
the fibers were allocated to the 849 quasars, 70 standard stars, a luminous red galaxy (required by the BOSS targeting algorithm), and 80 sky positions.
A typical epoch consists of eight to ten 15-minute exposures, totaling 65 hours when summed over the 32 epochs. 
The data were first processed through the BOSS spectroscopic pipeline \texttt{idlspec2d v5\_7\_1}, which performs flat-fielding, 1d extraction,
wavelength calibration, sky subtraction, and flux calibration \citep[][and references therein]{bolton12}.
The spectrum was resampled to a pixel scale of 69 km s$^{-1}$, which corresponds to roughly a half of the resolution element.
The flux calibration is tied to the PSF magnitudes of standard stars in the plate.
These pipeline-processed data were released to the public in the SDSS data release (DR) 12 \citep{alam15}.
The spectrophotometry was further improved with a custom flux-calibration algorithm; SDSS-RM observed 3.5 times more standard stars per plate
than the main BOSS survey, which enabled us to trace the spatial variation of the calibration more precisely \citep{shen14}.
Later in this work, we apply an additional aperture correction so that the decomposed host magnitude represents the total galaxy light.
The data of all epochs were stacked by inverse-variance weighted mean to create high SNR spectra, which are used throughout this work.
The SNR of the stacked spectra range from $\sim$5 to $\sim$300 per pixel, with the median value of $\sim$30.
The pixel flux errors were taken from the BOSS pipeline and properly propagated throughout the post-processing.
We corrected the spectra for Galactic dust extinction with the extinction curve of \citet{pei92} and the color excess $E_{\rm B-V}$ taken from \citet{schlegel98}.

\section{Data analysis \label{sec:analysis}}

We describe the details of the data analysis in this section.
In \S \ref{sec:methods}, we introduce our spectral decomposition technique which separates the galaxy component from an observed quasar spectrum.
This method is applied to the SDSS-RM data in \S \ref{sec:application}.
In \S \ref{sec:simulation}, we evaluate the reliability of the present method by Monte Carlo simulations and by comparisons with previous measurements.


\clearpage

\subsection{Spectral decomposition technique \label{sec:methods}}

We aim to extract host-galaxy information from the observed spectra with as little ambiguity as possible.
This is a challenge, since different combinations of quasar and stellar spectral models give rise to similar total spectra at UV-to-optical wavelengths.
For the present analysis, we choose the rest-frame spectral window from $\lambda$ = 3700 \AA\ to 5400 \AA, which includes a wealth of stellar absorption features
as well as important nebular lines such as [\ion{O}{2}] $\lambda$3727, H$\beta$, and [\ion{O}{3}] $\lambda\lambda$4959, 5007.
We devise a method to apply a series of spectral fits to the data 
in such a way that degeneracies between model components are minimized.
The following models are used to represent the relevant emission components, namely, the quasar accretion disk, gas in the broad line region (BLR),
narrow line region (NLR), and interstellar medium (ISM), and stars in the host galaxy.

\begin{enumerate}
\item Accretion disk: the emission from the quasar accretion disk can be represented by a single power-law continuum ($f_\lambda \propto \lambda^{\alpha_{\rm pl}}$)
in our spectral window, with the amplitude and slope being free parameters.
Previous observations have found slopes spanning the range of $-2.5 < \alpha_{\rm pl} < 0.0$, 
with a typical value of $\alpha_{\rm pl} \simeq -1.5$ \citep{richstone80, francis96, vandenberk01, ivezic02, pentericci03}.

\item Gas in the BLR, NLR, and ISM: 

\begin{enumerate}

\item 3700 \AA\ $< \lambda <$ 4450 \AA:
numerous high-order Balmer lines and metal lines as well as the Balmer continuum make up a pseudo-continuum in this spectral region.
It is modeled by subtracting the power-law contribution from the SDSS quasar composite spectrum provided by \citet{shen11}.
We take the composite of the highest-luminosity bin ($L_{5100} = 10^{45.5}$ erg s$^{-1}$) so that the host contamination is minimized;
\citet{shen11} report that the contamination at $\lambda =$ 5100 \AA\ is negligible in a quasar with $L_{5100} > 10^{45.0}$ erg s$^{-1}$.
The power-law contribution is determined in two continuum windows (in which strong emission lines are absent) at $\lambda$ = 4200 -- 4230 and 5470 -- 5500 \AA.
This pseudo-continuum template also includes individual BLR, NLR, and ISM lines present in the original composite spectrum, although the
ISM contribution should be minimal at this highest quasar luminosity.
We consider seven additional narrow lines or line complexes, namely, [\ion{O}{2}] $\lambda$3727, [\ion{Fe}{7}] $\lambda$3760, [\ion{Ne}{3}] $\lambda$3870, 
\ion{He}{1} $+$ H8 $\lambda$3890, [\ion{Ne}{3}] $\lambda$3969, H$\delta$, and H$\gamma$, to fit the ISM lines. 
They are each modeled with Gaussian profiles sharing the same velocity dispersion ($\sigma_{\rm g}$) and velocity offset relative to the object's formal redshift 
($v^{\rm off}_{\rm g}$), both of which are free parameters.

\item 4450 \AA\ $< \lambda <$ 5400 \AA:
the dominant contributors to this spectral region are the \ion{Fe}{2} pseudo-continuum, H$\beta$, and [\ion{O}{3}] $\lambda$4959, $\lambda$5007 lines.
We use the empirical \ion{Fe}{2} template of \citet{veron-cetty04} created from an observed spectrum of I Zw 1, a prototype narrow-line Seyfert 1 galaxy 
\citep[e.g.,][]{osterbrock85}.
This template is velocity-broadened with a Gaussian kernel with $\sigma_{\rm Fe II} = 1,500$ km s$^{-1}$; we test a different value of $\sigma_{\rm Fe II}$ later.
H$\beta$ is modeled with three Gaussian emission profiles with variable amplitudes, widths, and velocity offsets, while the two [\ion{O}{3}] lines 
are each modeled with two Gaussians.
The widths of these lines are varied independently from those of the above ISM lines.
We assume that the two [\ion{O}{3}] lines have the theoretical intensity ratio $\lambda$5007/$\lambda$4959 = 3.0 and that the two lines share the same profile
(i.e., widths and velocity offsets).

\end{enumerate}

\item Stars: we use simple stellar population (SSP) models of \citet{maraston11} to represent the stellar emission.
SSP is a single generation of stars formed in an instantaneous starburst; it is the simplest assumption for a stellar population and is suitable for the present analysis,
in which the dominant quasar light makes it difficult to exploit full details of the stellar properties.
We test more complicated stellar population models in \S \ref{sec:simulation}.
For SSP models, we adopt the STELIB \citep{leborgne03} stellar spectral library, the \citet{chabrier03} initial mass function (IMF), and solar metallicity.
Stellar age ($t_*$) takes values from 30 Myr to the age of the Universe at the redshift of each object.
The SSP spectra are velocity-broadened in accordance with the stellar velocity dispersion ($\sigma_*$) and the instrumental resolution\footnote{
The actual instrumental resolution varies as a function of wavelength, but here we assume the fixed mean value \citep[see][]{smee13}.}
assumed to be 
$\sigma_{\rm inst} = 65$ km s$^{-1}$,
and offset in wavelength by $v^{\rm off}_*$ relative to the object's redshift; both $\sigma_*$ and $v^{\rm off}_*$ are free parameters.
The amplitude of the SSP model determines the stellar mass $M_*$.
We test different choices of stellar library, IMF, and metallicity in \S 4.

\end{enumerate}

Our default models do not incorporate the effect of dust extinction. 
In fact, the amount of extinction observed in optically-selected broad-line quasars is usually quite small 
\citep{richards03,hopkins04,salvato09,matute12} and its effect is limited.
Recently \citet{krawczyk14} investigated a large sample of SDSS broad-line quasars and found that the vast majority 
are consistent with $E_{B-V} < 0.1$ mag.
Here we assume no dust extinction for both quasar nuclei and host galaxies, and later confirm that non-zero $E_{B-V}$ values
have little impact on our qualitative conclusions.
We note that there are dust-reddened broad-line quasars which can be identified in the near-infrared, although their surface density is much lower 
than that of optically-selected quasars \citep[e.g.,][]{glikman12}.
\citet{urrutia08} presented an enhanced merger fraction in dust-reddened quasars, hence their host stellar properties may be quite different from those
of normal optically-selected quasars.

\begin{figure*}
\epsscale{.9}
\plotone{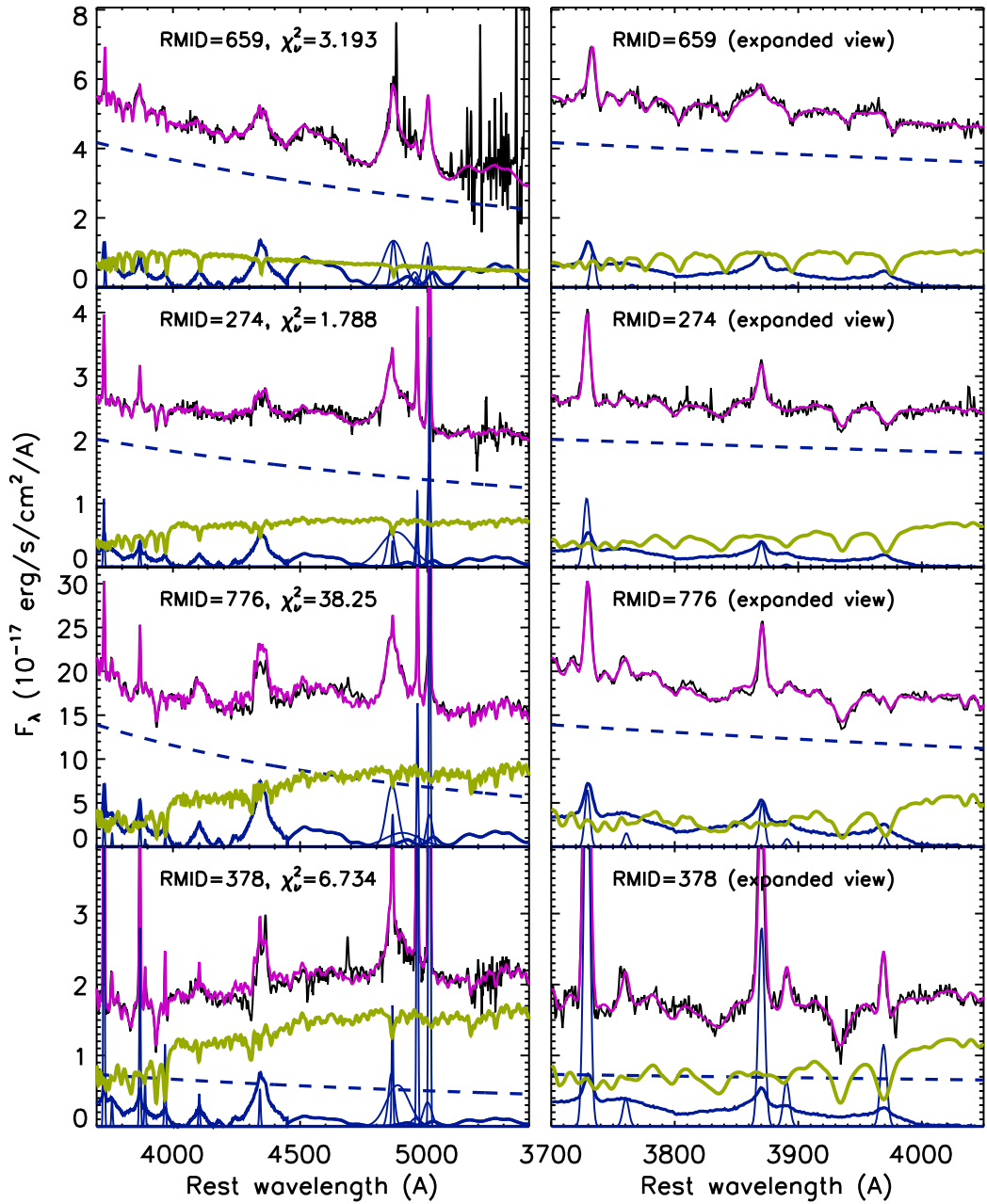}
\caption{Examples of the spectral decomposition. 
Each row corresponds to a single quasar in the full spectral window (left) and expanded view (right).
The first, second, and third rows show objects fitted with young ($t_*$ = 30 Myr), intermediate (1 Gyr), and old (8 Gyr) SSP models, respectively,
while the last row displays an object with strong narrow ISM emission lines.
Black lines indicate the observed co-added spectra, while the red lines show the total model spectra.
The other lines represent the individual model components, namely, the quasar power-law continuum (blue dashed),
the gas emission from the BLR, NLR, and ISM (blue solid), and the stellar emission (green).
\label{fig:spec_decomp}
}
\end{figure*}

We perform a series of spectral fits rather than fitting all the model components simultaneously. 
First, we determine the stellar age $t_*$ by fitting the spectrum over the rest-frame wavelength range $\lambda$ = 3700 -- 4050 \AA\ 
with the quasar power-law slope fixed to $\alpha_{\rm pl} = -1.5$.
This spectral window is chosen because 
(i) a wealth of strong stellar absorption features are present, including the Ca HK lines and the 4000 \AA\ break;
(ii) the quasar emission is relatively featureless;
(iii) the precise value of $\alpha_{\rm pl}$ is not important, since the flux ratio between the two edges 
of the window ($f_{4050}/f_{3700}$) changes only by $\sim$10 \% when $\alpha_{\rm pl}$ varies from $-1.5$ to the extreme values of $0.0$ or $-2.5$.
Second, the wider spectral range of $\lambda = $3700 -- 4750 \AA\ and 5050 -- 5400 \AA\ (i.e., the full 3700 -- 5400 \AA\ spectral window 
excluding the H$\beta$ region) is fit by varying 
all parameters but $t_*$, which is fixed to the value determined above.
We then return to the first step and update $t_*$ with the latest $\alpha_{\rm pl}$.
These two steps are iterated until $t_*$ converges, but in most cases the convergence occurs with no more than two iterations.
Finally, we subtract the best-fit model from the full spectrum and fit the residual with the H$\beta$ and [\ion{O}{3}] line models in the range $\lambda$ = 4750 -- 5050 \AA.

\clearpage
\subsection{Application to the data \label{sec:application}}

We apply the above algorithm to all the 191 SDSS-RM quasars found at $z < 1$; our spectral window ($\lambda$ = 3700 -- 5400 \AA) is shifted
beyond the spectrograph coverage at higher redshifts.
The standard $\chi^2$ method is used for the model fitting with the IDL routine {\tt MPFIT} \citep{markwardt09}.
As described above, the pixel flux errors were taken from the BOSS spectroscopic pipeline and propagated appropriately through the co-addition process.
Figure \ref{fig:spec_decomp} shows four typical examples of the decomposition results.
While the spectra are usually dominated by quasar light, most of the small-scale features are fit by the stellar models.
The median reduced $\chi^2$ of the fits over the whole spectral window is about five; as seen in Figure \ref{fig:spec_decomp}, this rather large value is
caused by the systematic discrepancy at certain wavelengths between the observed spectra and the models, the latter being not complex enough 
to provide perfect fits to every emission and absorption feature across the spectrum.
The uncertainties in the best-fit parameters are estimated following the Monte Carlo method outlined in \citet{shen11}.
We create 50 mock spectra by adding random noise to the original spectrum, using the Gaussian probability density function (PDF) with the standard 
deviation equal to the flux error at each pixel, and process them through the decomposition algorithm.
The measurement errors are evaluated from the 68 \% central intervals of the parameter distributions of the 50 trials.
We add additional 20 \% error to the derived stellar age, which accounts for the minimum uncertainty arising from non-continuous coverage of ages in the SSP models.
We perform further assessment of possible systematic uncertainties in the next section.

\begin{figure}
\epsscale{1.1}
\plotone{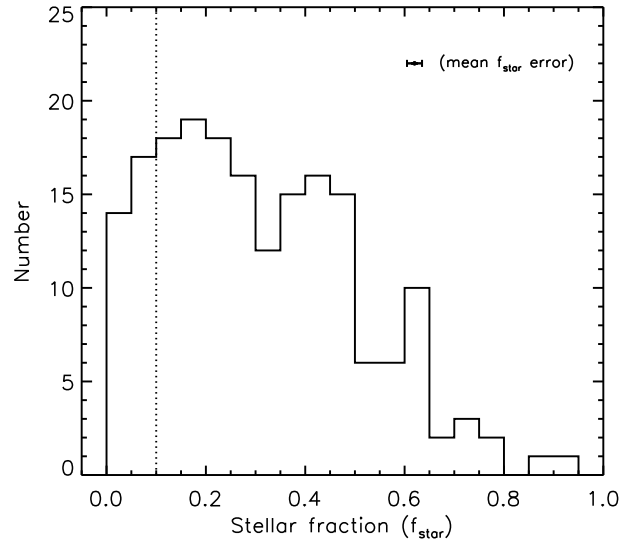}
\caption{Histogram of the fractional stellar contribution to the total flux ($f_*$), measured at $\lambda$ = 4000 \AA\ in the rest frame, 
for the 191 SDSS-RM quasars at $z < 1$.
The typical $f_*$ uncertainty is shown with the error bar at the top right corner.
We deem the decomposition analysis successful for 156 objects (82 \% of the initial 191 quasars) with $f_* > 0.1$ (dotted line).
\label{fig:Fssp}
}
\end{figure}

Figure \ref{fig:Fssp} presents the histogram of the fractional contribution of the stellar component to the total flux ($f_*$) measured at $\lambda$ = 4000 \AA\ 
in the rest frame.
We exclude the 31 objects with $f_* < 0.1$ from the following analysis; we will demonstrate in the next section that their host results are unreliable.
An additional four objects are excluded because they have particularly large statistical uncertainties in the host stellar mass, $\Delta$log($M_*/M_\odot) >$ 0.5 dex.
These excluded objects tend to have high quasar luminosities, as shown in Figure \ref{fig:zMi}.
Our final sample consists of the remaining 156 quasars, more than 80 \% of the initial 191 objects.
The median redshift and absolute magnitude of the 156 quasars are ${\langle}z{\rangle} = 0.72$ and ${\langle}M_i{\rangle} = -22.2$ mag, respectively.



We focus on six physical quantities in this work, namely, stellar age $t_*$, stellar mass $M_*$, SFR, stellar velocity dispersion $\sigma_*$,
[\ion{O}{3}] $\lambda$5007 luminosity $L_{\rm [OIII]}$, and SMBH mass $M_\bullet$.
The parameters $t_*$, $M_*$, $\sigma_*$, and $L_{\rm [OIII]}$ are taken directly from the decomposition results.
Since the quantity $\sigma_*$ is poorly constrained in many cases, its measurement is rejected if the best-fit value lies outside the physically plausible range 
($10 < \sigma_* < 400$ km s$^{-1}$) or smaller than three times the estimated statistical error.
SFRs are estimated from the [\ion{O}{2}] $\lambda$3727 luminosity.  
As shown in Figure \ref{fig:spec_decomp}, the [\ion{O}{2}] line is comprised of two components, i.e., the quasar BLR/NLR template and the host ISM line with Gaussian profile.
The former contribution is tied to the overall amplitude of the BLR/NLR template, which is fit over the full spectral window.
We use the luminosity of the latter ISM component ($L'_{\rm [OII]}$) for SFR estimates, assuming the \citet{kennicutt98} calibration:
\begin{equation}
{\rm SFR}\ (M_\odot\ {\rm yr}^{-1})\ =\ (1.4 \pm 0.4) \times 10^{-41}\ L'_{\rm [O II]}\ {\rm (erg\ s}^{-1}{\rm )} .
\label{eq:sfr}
\end{equation}
Since the NLR usually contributes only weakly to this low-ionization line, it may be a good tracer of star formation in quasar host galaxies 
even without spectral decomposition \citep{ho05}.
On the other hand, \citet{kim06} found that the [\ion{O}{2}]/[\ion{O}{3}] ratios observed in SDSS AGNs at $z < 0.3$ 
are fully consistent with
AGN photoionization, hence there is no need to invoke any additional [\ion{O}{2}] source such as star formation.
Although the mean quasar contribution is properly subtracted with the BLR/NLR template in this work, we conservatively treat the derived SFRs as rough estimates only,
given that our single BLR/NLR template cannot trace object-to-object variation of the [\ion{O}{2}] strength. 
Another concern is the effect of dust extinction, which we discuss later.


\begin{figure*}
\epsscale{.9}
\plotone{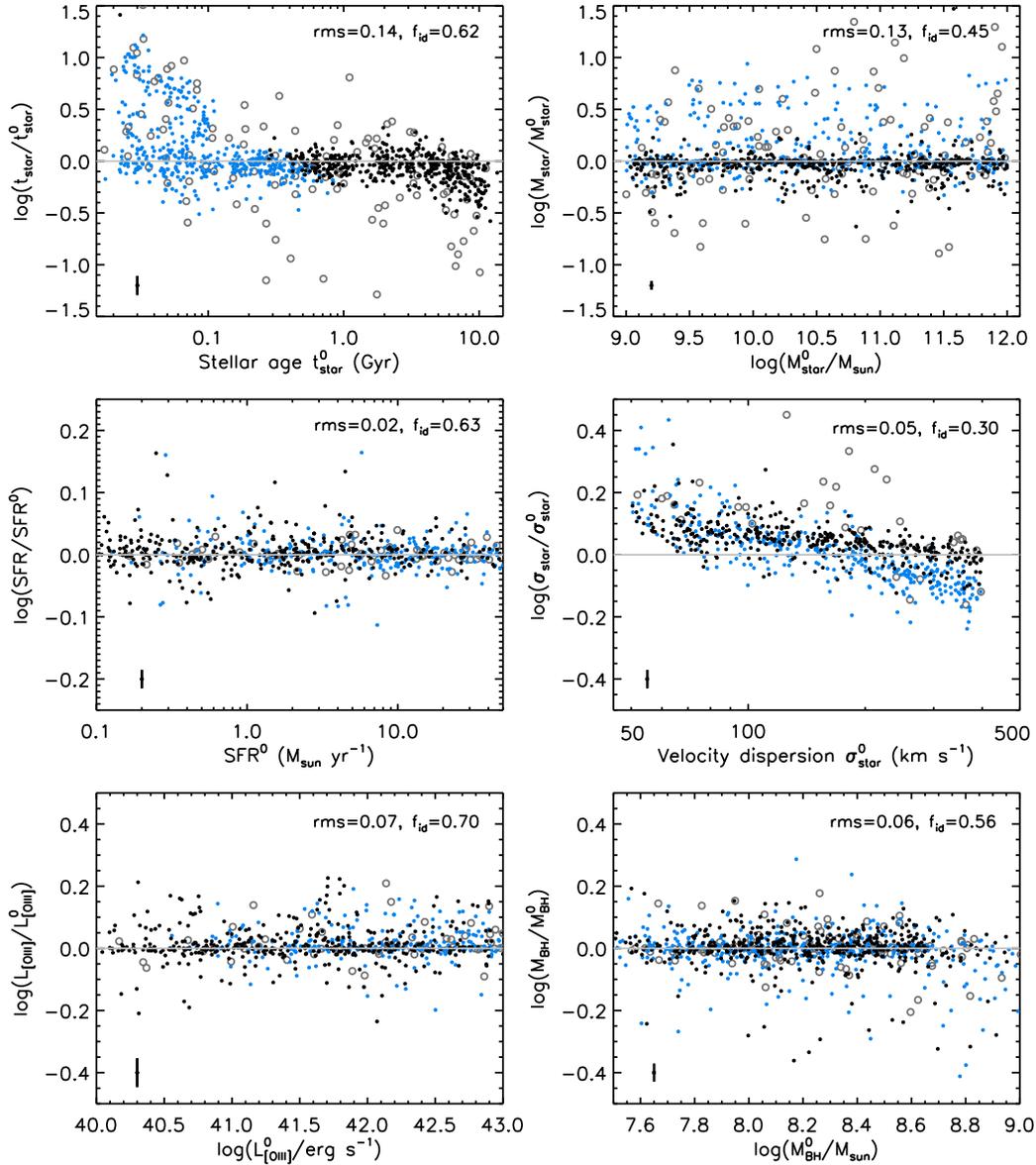}
\caption{Deviations of the measured (output) values from the intrinsic (input; the quantities with the superscript ``0") values for $t_*$ (top left), $M_*$ (top right), 
SFR (middle left), $\sigma_*$ (middle right), $L_{\rm [OIII]}$ (bottom left), and $M_\bullet$ (bottom right)
for simulated spectra. 
The black dots represent the spectra with reliable fits of the stellar component (output stellar fraction $f_* > 0.1$, mass error $\Delta$log $M_* < 0.5$ dex, 
and stellar age $t_* > 0.5$ Gyr), while the blue dots represent those with younger stellar ages ($f_* > 0.1$, $\Delta$log $M_* < 0.5$ dex, and $t_* < 0.5$ Gyr;
we find few such cases in the real quasars as described in the text).
The open circles show the remaining spectra with poor fits of the stellar component ($f_* < 0.1$ or $\Delta$log $M_* > 0.5$ dex).
The symbols in the upper left panel are given small random offsets in stellar age to improve visibility.
The typical statistical error is shown by the error bar at the bottom left corner of each panel.
The horizontal lines represent the identity lines.
The numbers in each panel indicate, for the black dots, the RMS scatter around the identity line and the fraction ($f_{\rm id}$) of those whose
ordinate values are consistent with zero within 1$\sigma$ statistical error.
\label{fig:simulation}
}
\end{figure*}

SMBH masses are estimated with 
the single-epoch virial estimator \citep{vestergaard06}:
\begin{eqnarray}
{\rm log} \left(\frac{M_{\rm BH}}{M_\odot} \right) & = & {\rm log} \Biggl( \Biggl[ \frac{{\rm FWHM (H\beta)}}{1000\ {\rm km\ s^{-1}}} \Biggr]^2 
\Biggl[ \frac{{\lambda L_\lambda (5100\ {\rm \AA})}}{10^{44}\ {\rm erg\ s^{-1}}} \Biggr]^{0.5} \Biggr) \nonumber \\
& & + (6.91 \pm 0.02).
\label{eq:mbh}
\end{eqnarray}
It is believed that this H$\beta$-based estimator is more reliable than other single-epoch estimators, such as those based on \ion{C}{4} $\lambda$1549 or 
\ion{Mg}{2} $\lambda$ 2800 \citep{shen12, shen13}.
In some cases our H$\beta$ models apparently fit the underlying continuum residuals rather than the line itself, especially when the SNR around H$\beta$ is low.
The $M_\bullet$ measurements were rejected for 16 such fits identified by visual inspection.
We will eventually obtain more precise SMBH masses for all the objects with RM analysis.


\begin{figure*}
\epsscale{1.1}
\plotone{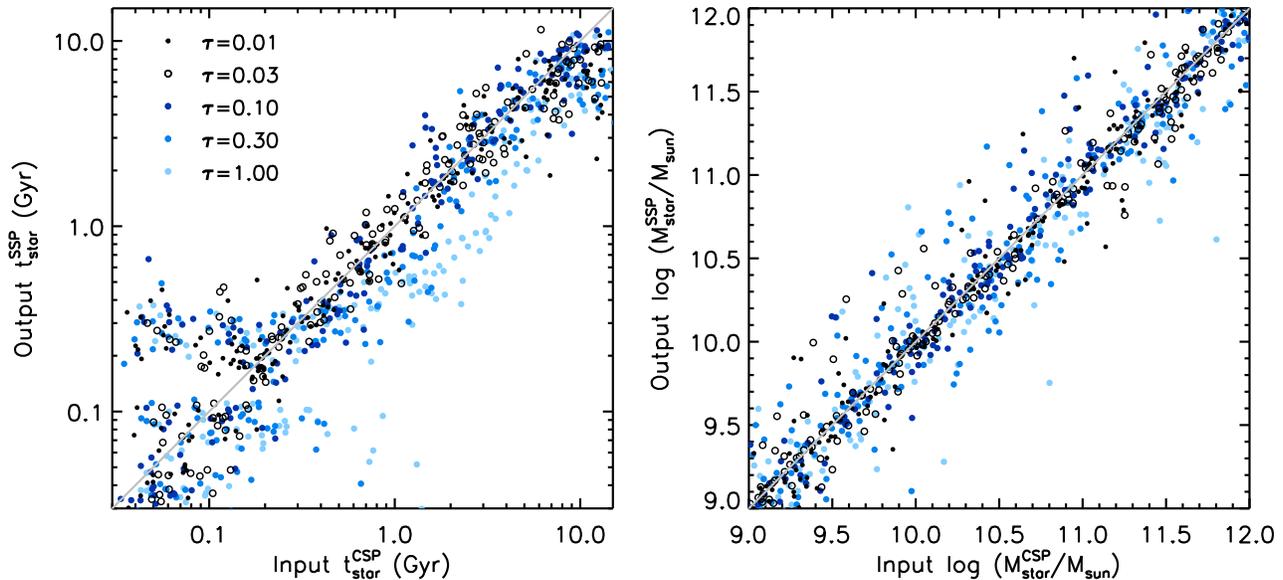}
\caption{Comparison between the input CSP maximum age $t_*^{\rm CSP}$ and the output SSP age $t_*^{\rm SSP}$ (left) or between
the accumulated CSP mass $M_*^{\rm CSP}$ and the SSP mass $M_*^{\rm SSP}$ (right).
The different symbols represent different values of the star formation timescale $\tau$ (in units of Gyr) as reported in the left panel.
The points in the left panel are given small random offsets to improve visibility.
The gap at $t_*^{\rm SSP} \sim 0.15$ Gyr is caused by the sparse time sampling of our SSP models around this age.
The diagonal lines represent the identity lines.
The quantities $t_*^{\rm SSP}$ and $t_*^{\rm CSP}$ are largely consistent at $t_*^{\rm CSP} > 0.1$ Gyr and $\tau <$ a few 100 Myr, while
$t_*^{\rm SSP}$ becomes smaller at larger $\tau$ due to the prolonged star formation.
There is little systematic offset between the input and output masses.
\label{fig:simulation_csp}
}
\end{figure*}

The limited aperture size of the BOSS spectrograph fibers does not include all the light from the extended hosts.
The fiber diameter of 2\arcsec\ corresponds to approximately 14 kpc at the median redshift of our sample ($z = 0.72$).
We correct the stellar mass, derived from the decomposed galaxy flux ($f_{\rm g}^{\rm fiber}$), for the aperture loss as follows.
Since the spectrophotometry of the BOSS spectroscopic pipeline is tied to the PSF magnitude of standard stars, the quasar flux decomposed from each spectrum
($f_{\rm q}$) should correctly represent the total flux of the unresolved quasar nucleus.
The SDSS imaging pipeline measures the total nuclear plus host flux ($f_{\rm tot}$) 
with the \texttt{cModel} algorithm \citep{abazajian04}, which fits galaxy models to the observed image profile.
Therefore, the total host flux ($f_{\rm g}$) is given by $f_{\rm g} = f_{\rm tot} - f_{\rm q}$.
Multiplying the stellar mass of each source measured from the spectra by the ratio $f_{\rm g}/f_{\rm g}^{\rm fiber}$
provides a crude estimate
of the total stellar mass, with the implicit assumptions that the mass-to-luminosity ratio in the fiber represents that in the whole galaxy
and that quasar variability between the epochs of imaging (before 2005) and spectroscopic (2014) observations does not introduce a significant bias.
This aperture correction is calculated in the observed-frame $r$ band.
The correction is not applied to the SFRs, since broad-band magnitudes are a poor tracer of the spatial extent of the [\ion{O}{2}] line emission.
When we later discuss the relation between SFR and stellar mass, we use the stellar mass within the fiber ($M_*^{\rm fiber}$;
the stellar mass before the aperture correction) so that the two quantities refer to the same central part of the galaxies.
Therefore, the present analysis does not map SFR and its relation to stellar mass to extended regions of the galaxies outside the fibers.



\subsection{Reliability of the present method \label{sec:simulation}}

Here the reliability of the present method is evaluated with Monte Carlo simulations.
We generate 1,000 mock spectra by combining spectral models of quasar accretion disk, gas in the BLR, NLR, ISM, and host stellar population 
as described in \S \ref{sec:methods}, with realistic PDFs for the parameter values.
The power-law slope $\alpha_{\rm pl}$ is assumed to follow a Gaussian PDF with mean $-1.5$ and standard deviation $0.3$,
which approximately reproduces past measurements \citep{ivezic02, pentericci03}.
The SSP age is randomly drawn from 14 logarithmically spaced values between $t_* = $ 30 Myr and 10 Gyr, while the stellar mass is distributed uniformly on 
a logarithmic scale between log ($M_*/M_\odot$) = 9 and 12.
The ratio between the mean amplitudes of the stellar model and the quasar power law is assumed to follow a logarithmically-uniform PDF between 0.01 and 10.0.
The stellar velocity dispersion is drawn from a logarithmically-uniform PDF between 50 and 400 km s$^{-1}$.
All other model parameters are assumed to follow the Gaussian PDFs best describing the distributions derived in our own measurements. 
For each of the mock spectra, we randomly select a SDSS-RM quasar at $z < 1$ and apply its redshift and pixel flux errors to mimic a real spectrum.

These mock spectra are processed through the decomposition algorithm used for the real data.
Figure \ref{fig:simulation} presents the comparisons between the intrinsic (input) and measured (output) parameter values. 
The stellar age and mass show relatively large scatter, reflecting the ambiguity in decomposing the nuclear and stellar components 
from the observed spectra.
The scatter is significant at $t_*^0 \la 0.1$ Gyr (the superscript ``0" represents the input parameter), where the relatively blue spectra of 
young stars are difficult to distinguish from those of quasar nuclei.
These degeneracies result in over-estimated ages up to $t_* \sim 0.5$ Gyr, hence 
we conservatively regard measured stellar ages at $t_* \le 0.5$ Gyr as upper limits (although we find only few such cases in the real quasars; see below).
At larger ages, most of the scatter is caused by spectra with $f_* < 0.1$ or $\Delta$ log $M_* > $ 0.5 dex. 
When these simulated quasars are eliminated, as we did for the real sample, all six parameters are constrained within a scatter of roughly 0.15 dex;
the actual RMS scatter of each parameter is reported in each panel of the figure.
Figure \ref{fig:simulation} also reports for each parameter the fraction ($f_{\rm id}$) of the simulated spectra whose output value is consistent with the input 
(i.e., the ordinate value is consistent with zero) within 1$\sigma$ statistical error.
The fraction for $\sigma_*$ is low, due to the systematic over-estimation of this parameter at small $\sigma_*^0$ where the measurement is
most difficult.
For the other five parameters, $f_{\rm id}$ are somewhat smaller than but roughly comparable to the expected 68 \%.

\begin{figure*}
\epsscale{1.1}
\plotone{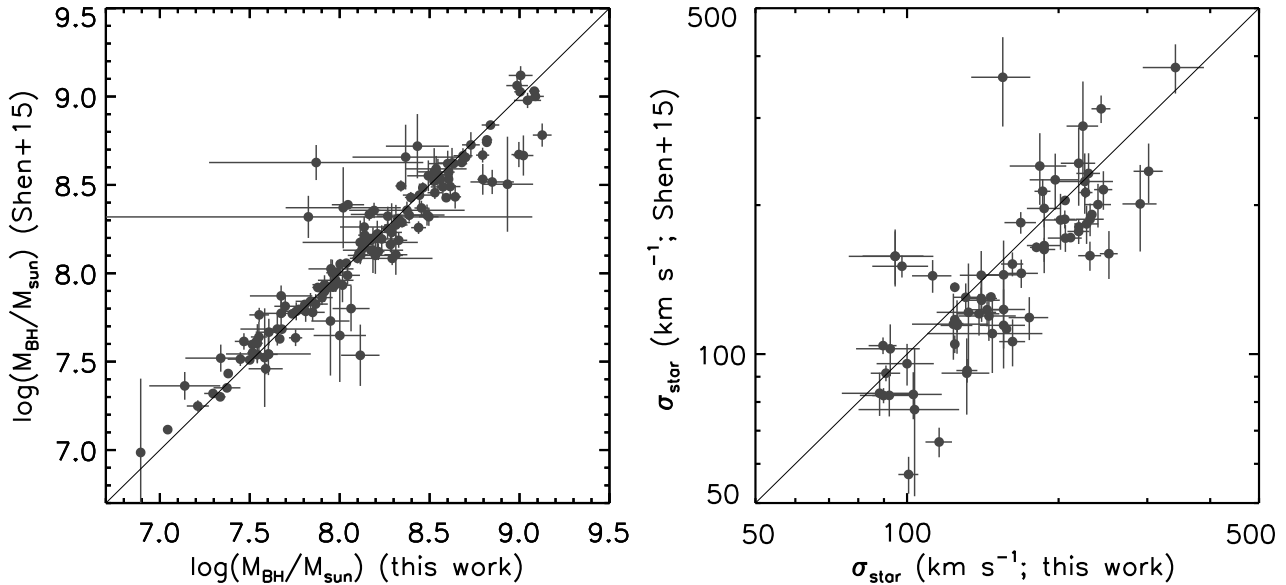}
\caption{Comparison of $M_\bullet$ (left; 140 successful measurements) or $\sigma_*$ (right; 73 successful measurements)
for common objects measured by \citet{shen15} and this work. 
The diagonal lines represent the identity lines.
The two sets of measurements are in good agreement with each other, although there is a small ($\sim$10 km s$^{-1}$) systematic offset in $\sigma_*$.
\label{fig:shen15}
}
\end{figure*}

We also perform a set of simulations that replace the input SSP models with composite stellar population (CSP) models with exponentially declining 
star formation history.
CSP spectra are calculated as 
\begin{equation}
f_\lambda (t_*^{\rm CSP}) = \int_{0}^{t_*^{\rm CSP}} \psi (t_*^{\rm CSP} - t) f_\lambda^{\rm SSP} (t) dt ;\  \psi (t) = e^{-t/\tau} ,
\end{equation}
where $t_*^{\rm CSP}$ is the time since the onset of star formation (i.e., the maximum CSP age), $\psi (t)$ is the star formation history with 
the exponentially declining timescale $\tau$, and $f_\lambda^{\rm SSP} (t)$ is an SSP spectrum.
We create 1,000 mock spectra by combining these CSP models with other emission components as above, and process them through the decomposition 
algorithm (which uses the SSP models to represent stellar spectra).
The results are shown in Figure \ref{fig:simulation_csp} for stellar age and mass; the other parameters are less sensitive to the choice of star formation history.
The decomposition is unreliable for input $t_*^{\rm CSP} \la 0.1$ Gyr, which is similar to the situation described above.
At larger $t_*^{\rm CSP}$, the output SSP age is consistent with $t_*^{\rm CSP}$ when $\tau$ is less than a few 100 Myr,
while the former becomes systematically small at larger $\tau$ due to the prolonged star formation.
There is little systematic offset between the output SSP mass and the input CSP mass, the latter being the mass accumulated during the whole star formation history.

Finally, we compare our measurements of $M_\bullet$ and $\sigma_*$ with the previous measurements 
by \citet{shen15}, who investigated the $M_\bullet$ -- $\sigma_*$ relation of 
the SDSS-RM quasars at $z < 1$ using the same co-added spectra as in this work.
They performed spectral decomposition with the principal component analysis (PCA) method following \citet{vandenberk06}.
The PCA method has the advantage of being highly efficient in reproducing observed spectra with an optimal number of empirical eigenspectra,
but the interpretation of the reconstructed spectra is usually not straightforward.
The SMBH masses were derived with the \citet{vestergaard06} estimator as in this work, while
the stellar velocity dispersion was measured with the \texttt{vdispfit} routine in the \texttt{idlspec2d} package and with the penalized pixel-fitting code
\citep{cappellari04} in the spectral window $\lambda$ = 4125 -- 5350 \AA. 
Figure \ref{fig:shen15} compares the $M_\bullet$ and $\sigma_*$ values measured by the two studies.
They are in good agreement with each other, although there is a slight systematic offset in $\sigma_*$;
our measurements give $\sim$10 km s$^{-1}$ larger values than those of \citet{shen15} on average.
The root-mean-square (RMS) scatter around this offset is $\sim$ 40 km s$^{-1}$, while the typical measurement error is $\sim$15 km s$^{-1}$ in the both studies.\footnote{
Note that the two studies are not independent, since they use the same spectra for the $\sigma_*$ measurements.}
Our decomposition procedure is not optimized for the $\sigma_*$ measurements, since the spectral resolution of the adopted stellar library
($\sigma \sim 80$ km s$^{-1}$) is not sufficiently high for this analysis; we chose this library since it covers the whole of our spectral window 
with a wide range of stellar parameters, which is crucial for a detailed study of the decomposed stellar emission.
In contrast, the emphasis is on the most reliable $\sigma_*$ measurements in \citet{shen15}, who used two dedicated routines 
adopting spectral libraries with high resolution ($\sigma \le 25$ km s$^{-1}$).
Our fitting procedure also includes Ca H and K lines, which tend to over-estimate $\sigma_*$ \citep{kormendy82, bernardi03, greene06}.


\section{Results \label{sec:results}}

\begin{figure*}
\epsscale{.9}
\plotone{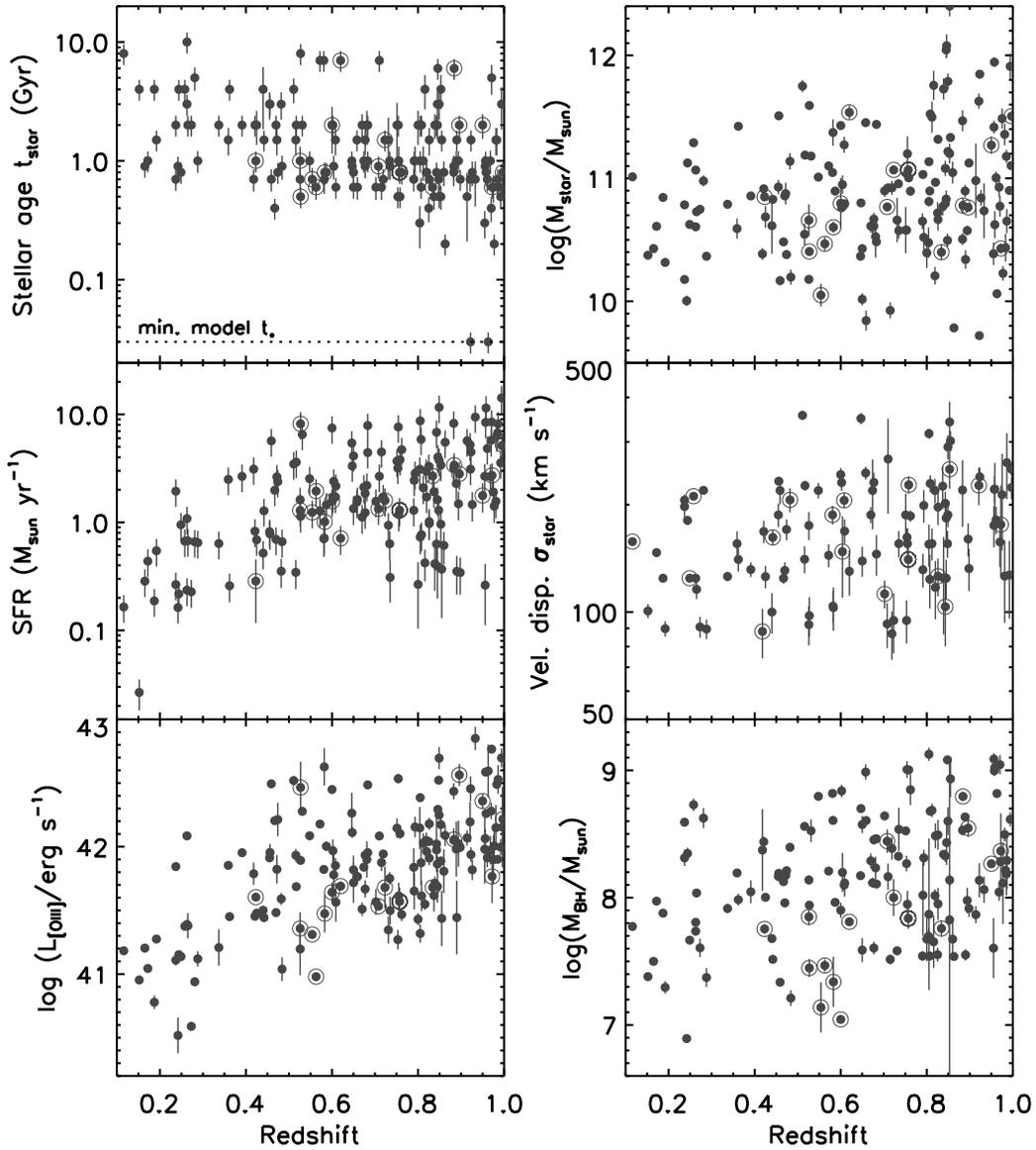}
\caption{Spectral decomposition results of our sample of 156 quasars with successful measurements.
Shown are $t_*$ (top left), $M_*$ (top right), SFR (middle left), $\sigma_*$ (middle right), $L_{\rm [OIII]}$ (bottom left), and $M_\bullet$ (bottom right), 
all as a function of redshift.
The objects with the stellar fraction $f_* > 0.5$ are marked with the large open circles.
The error bars indicate statistical errors; there may be additional systematic uncertainties as discussed in the text.
The dotted line in the top left panel represents the minimum age of our SSP models, $t_*$ = 30 Myr; the two objects on this line are well fitted with this age
but may have even younger stellar populations.
\label{fig:mainp3}
}
\end{figure*}
\begin{figure*}
\epsscale{1.1}
\plotone{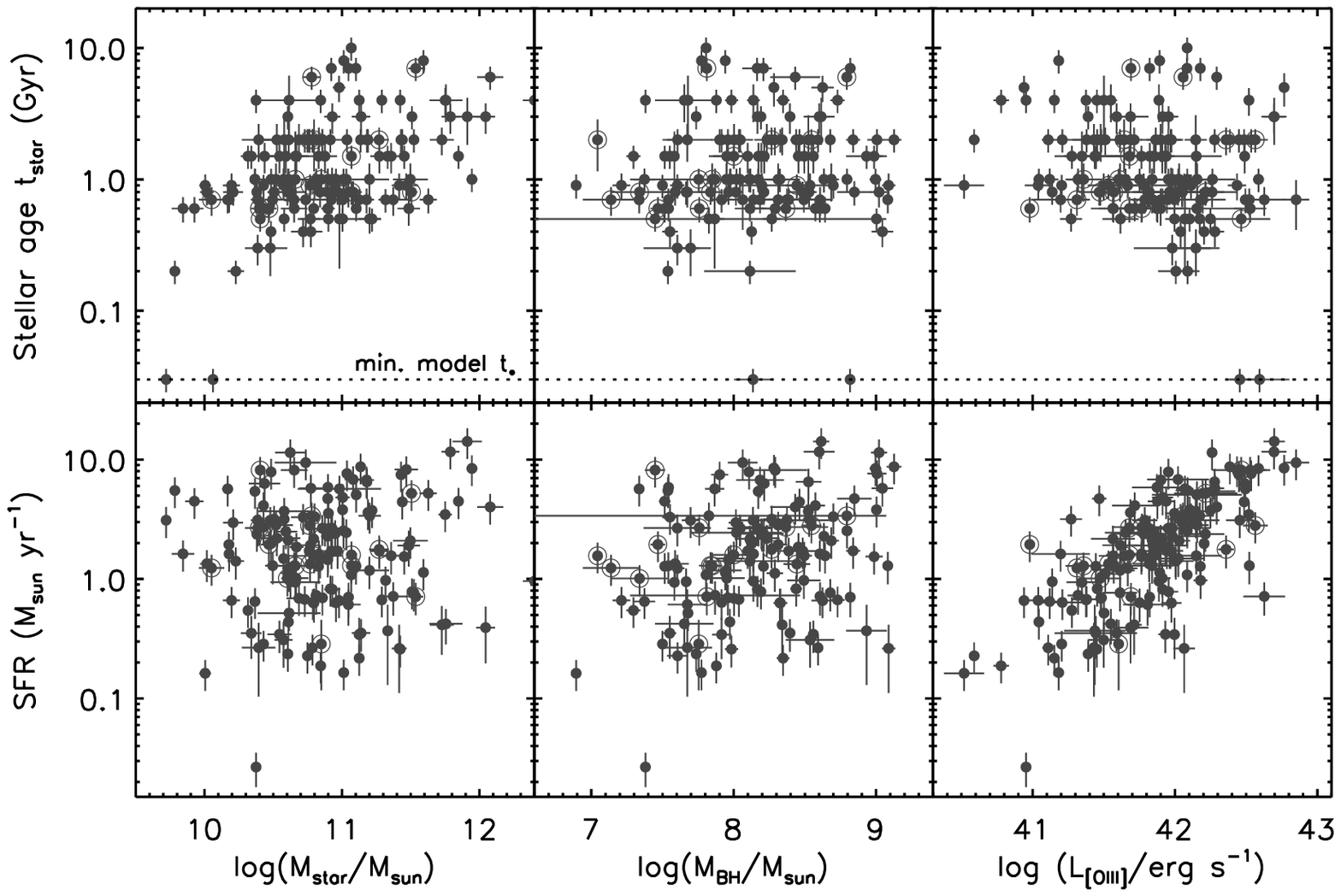}
\caption{Relations between $t_*$ (top panels) or SFR (bottom panels) and $M_*$ (left), $M_\bullet$ (middle), or $L_{\rm [OIII]}$ (right).
All the successful measurements of the 156 quasars are plotted.
The objects with the stellar fraction $f_* > 0.5$ are marked with the large open circles.
The dotted lines in the top panels represent the minimum age of our SSP models, $t_*$ = 30 Myr.
\label{fig:mainp}
}
\end{figure*}
\begin{figure}
\epsscale{1.1}
\plotone{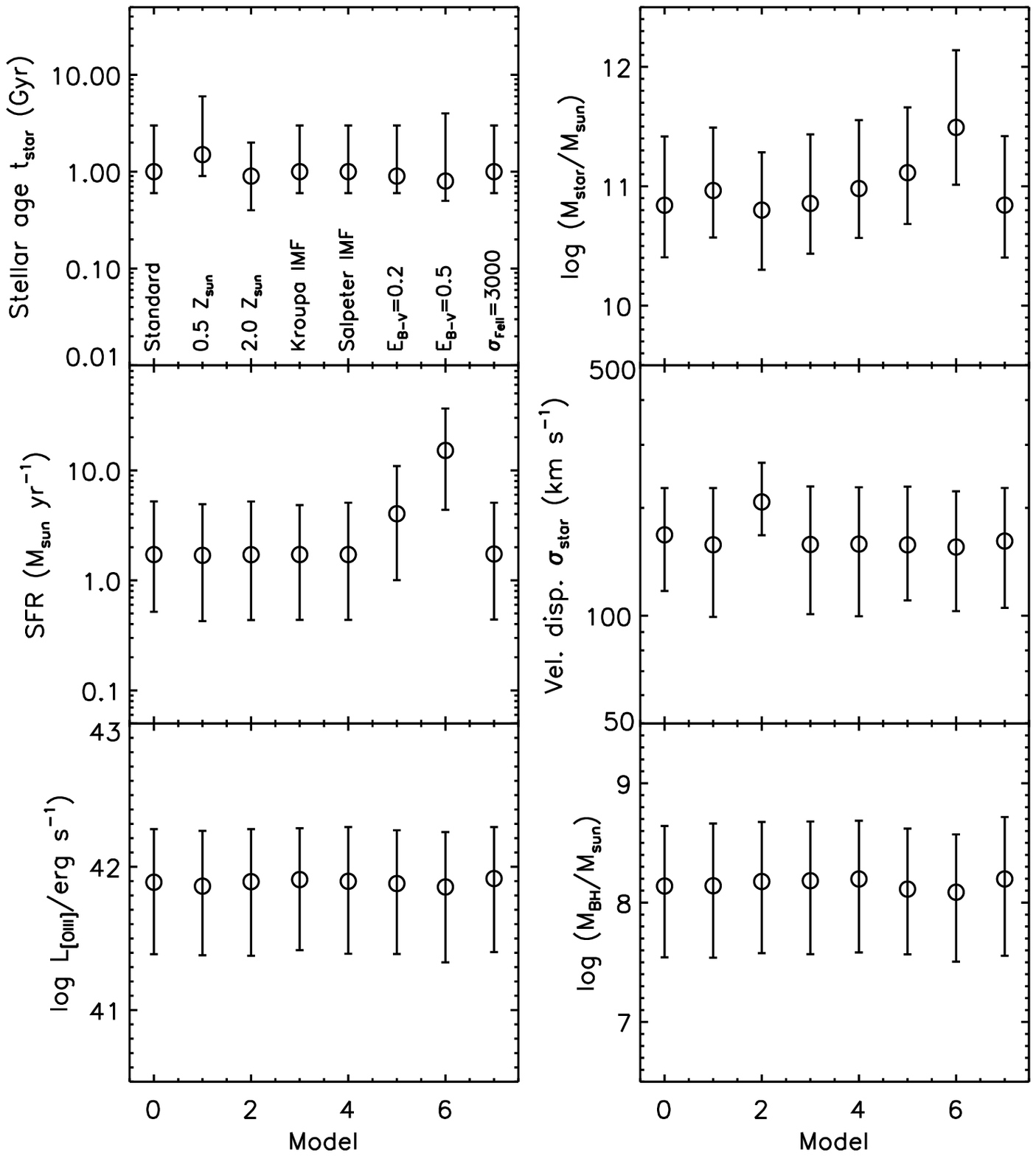}
\caption{Spectral decomposition results with different model specifics as summarized in Table \ref{tab:ssp_specifics}; 
Model 0 (standard), 1 ($Z = 0.5 Z_\odot$), 2 ($Z = 2.0 Z_\odot$),
3 (Kroupa IMF), 4 (Salpeter IMF), 5 ($E_{B-V} = 0.2$ mag), 6 ($E_{B-V} = 0.5$ mag), and 7 ($\sigma_{\rm Fe II}$ = 3000 km s$^{-1}$).
The median values (circles) and 68 \% central intervals (error bars) of the derived parameter distributions are shown for
$t_*$ (top left), $M_*$ (top right), SFR (middle left), $\sigma_*$ (middle right), $L_{\rm [OIII]}$ (bottom left), and $M_\bullet$ (bottom right).
\label{fig:ssp_specifics}
}
\end{figure}

We now return to the actual spectra.
The derived quasar and host properties are displayed
in several parameter planes in Figures \ref{fig:mainp3} and \ref{fig:mainp}.
Their characteristic values (medians and 68\% central intervals) are also presented in Figure \ref{fig:ssp_specifics} (``Model 0").
The stellar ages cluster around $t_*$ = 1 Gyr and show a clear deficit at less than several hundred Myr, suggesting that the galaxies have not experienced major star formation
in the recent past.
The absence of young stars is consistent with the relatively low SFRs (within the spectroscopic fibers), which are less than 10 $M_\odot$ yr$^{-1}$ in most cases.
Our results agree with those of \citet{ho05} and \citet{kim06}, who found that the SFRs in broad-line quasars at $z \le 1$ are $\sim$ 10 $M_\odot$ yr$^{-1}$ 
at most, and can be much smaller considering the contribution from AGN photoionization to lines like [\ion{O}{2}].
On the other hand, \citet{shi09} report moderately higher SFRs ($\sim 10 M_\odot$ yr$^{-1}$) than ours for a small number of SDSS quasars at $z \sim 1$, 
based on observations of polycyclic aromatic hydrocarbon emission.
At least a part of this discrepancy may be due to selection effects, as their sample is about an order of magnitude more luminous than the present sample.
We stress that these are optically-selected broad-line quasars; dust-reddened or obscured quasars are different populations (see \S 1) and
may have much higher SFRs than those estimated here.
The stellar and SMBH masses of the present sample are in the range of $10^{10} M_\odot < M_* < 10^{12} M_\odot$ 
and $10^{7} M_\odot < M_\bullet < 10^{9} M_\odot$, respectively, 
suggesting that the SDSS-RM quasars are hosted by fairly massive galaxies with massive SMBHs.
Finally, Figure \ref{fig:mainp3} demonstrates that objects with larger stellar fraction ($f_* > 0.5$) tend to have lower $L_{\rm [O III]}$ and $M_\bullet$ at a given redshift,
while their host properties are not significantly different from those with smaller $f_*$.


We checked the robustness of our analysis to the model specifics, using the seven different basic models summarized in Table \ref{tab:ssp_specifics}.
Models 1 and 2 assume the metallicity $Z = 0.5 Z_\odot$ and $2.0 Z_\odot$, respectively, while Models 3 and 4 adopt different IMFs from our default 
\citet{chabrier03} IMF.
In Models 5 and 6, the host galaxies (i.e., the stellar and ISM model components) are subject to modest dust extinction, using the Small Magellanic Cloud type 
extinction curve \citep[$R_\lambda = A_\lambda/E_{B-V} = 1.39\ (\lambda/10000)^{-1.2}$;][]{prevot84,richards03,hopkins04}.
We do not apply this additional dust extinction to the quasar component, since the adopted empirical quasar templates already include the mean reddening,
and broad-line quasars are usually observed with little extinction (see the discussion in \S \ref{sec:methods}).
Finally, Model 7 assumes twice the broadening width of the \ion{Fe}{2} template compared to the standard model.
The spectral decomposition results with these different models are displayed in Figure \ref{fig:ssp_specifics}.
The lower/higher metallicity leads to older/younger stellar ages, as expected from the well-known age-metallicity degeneracy, and slightly higher/lower
stellar masses.
The \citet{salpeter55} IMF results in $\sim$0.2 dex higher $M_*$ than the \citet{chabrier03} or \citet{kroupa01} IMF, which reflects higher mass-to-luminosity 
ratios by a similar factor.
Assuming a color excess of $E_{B-V}$ = 0.2 or 0.5 mag 
results in the stellar masses and SFRs simply increasing by the corresponding extinction factors ($A_V \sim$ 0.6 or 1.5 mag), while the other parameters remain 
nearly unchanged.
The assumption of the broadening width of the \ion{Fe}{2} template has little impact. 
Overall, our results are qualitatively insensitive to the specifics of the spectral models; the quasar host galaxies are characterized by intermediate stellar ages 
($t_* \sim 1$ Gyr), fairly large stellar masses ($M_* \sim 10^{11} M_\odot$), relatively low SFRs ($\la 10 M_\odot$ yr$^{-1}$ unless the dust extinction
is very large), and large SMBH masses ($M_\bullet \sim 10^{8} M_\odot$).

\begin{table}
\begin{center}
\caption{Model specifics. \label{tab:ssp_specifics}}
\begin{tabular}{ll}
\tableline\tableline
Model &  Description        \\
\tableline
 0\tablenotemark{a} ... & Standard\\     
 1 ... & $Z = 0.5 Z_\odot$\\     
 2\tablenotemark{b} ... & $Z = 2.0 Z_\odot$\\ 
 3 ... & \citet{kroupa01} IMF  \\
 4 ... & \citet{salpeter55} IMF\\
 5 ... & $E_{B-V} = 0.2$ mag    \\
 6 ... & $E_{B-V} = 0.5$ mag\\
 7 ... & $\sigma_{\rm Fe II}$ = 3,000 km s$^{-1}$  \\
\tableline
\end{tabular}
\tablenotetext{a}{The standard model adopts $Z = Z_\odot$, \citet{chabrier03} IMF, $E_{B-V} = 0.0$ mag, and $\sigma_{\rm Fe II}$ = 1,500 km s$^{-1}$.}
\tablenotetext{b}{The STELIB stellar library is replaced with MILES which has a wider coverage of stellar ages at this metallicity in the \citet{maraston11} models.}
\end{center}
\end{table}

\begin{figure}
\epsscale{1.1}
\plotone{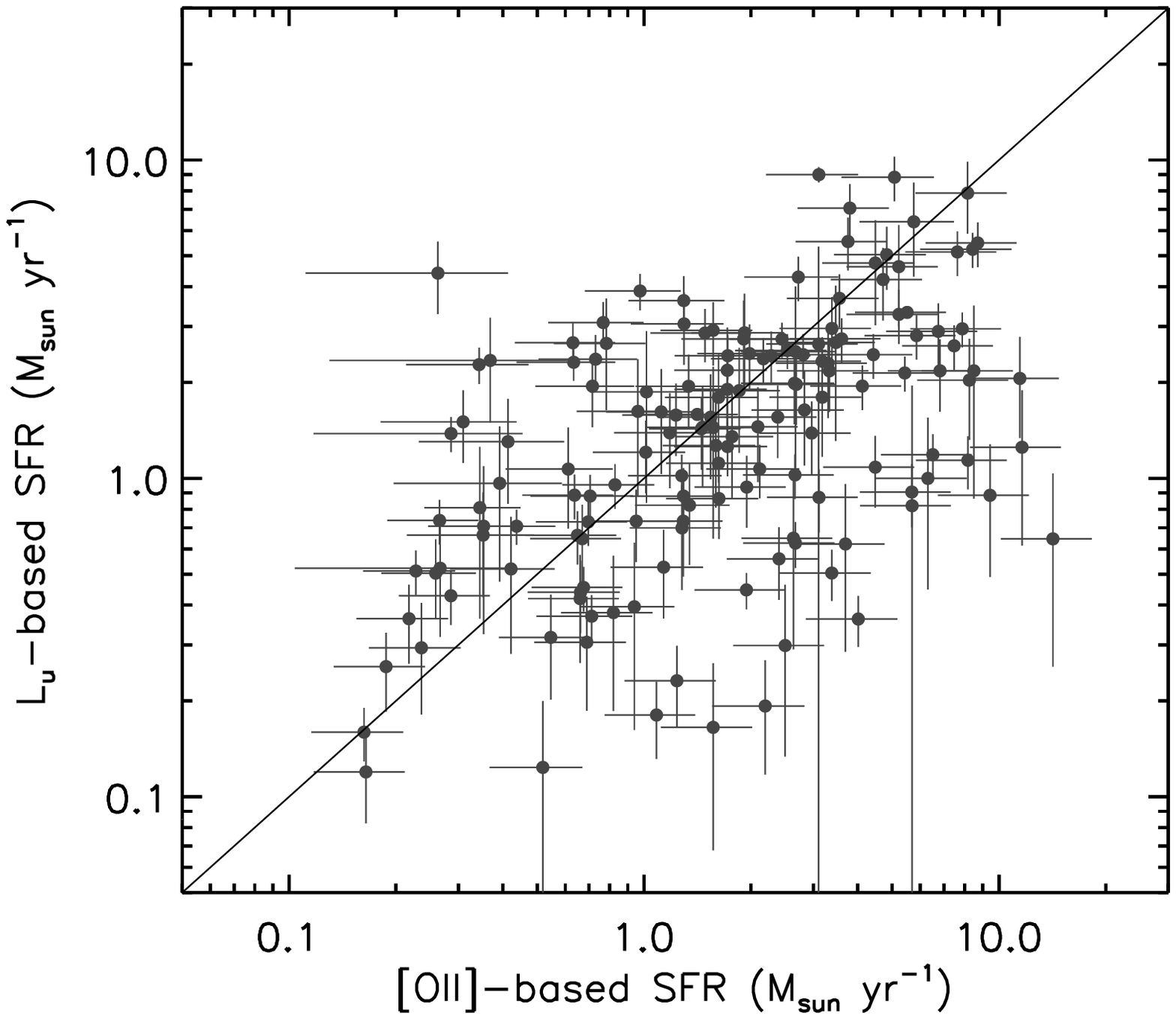}
\caption{Comparison between the $L_u$-based and [\ion{O}{2}]-based SFRs.
The diagonal solid line represents the identity line.
The two estimates are roughly comparable, although there is a large scatter.
\label{fig:sfrs}
}
\end{figure}

As an additional test of the reliability of the SFR estimates, we compare them with SFRs estimated with a different calibrator, using the rest-frame
$u$-band luminosity ($L_u$) of the stellar component.
We use the \citet{hopkins03} calibration to derive the $L_u$-based SFR:
\begin{equation}
   {\rm SFR}\ (M_\odot\ {\rm yr}^{-1}) = \Biggl( \frac{L_u}{1.81 \times 10^{21}\ {\rm W\ Hz^{-1}}} \Biggr)^{1.186} ,
\end{equation}
with $L_u$ calculated from the decomposed host spectrum of each object, extrapolated to $\lambda < 3700$ \AA\ (i.e., outside the fitting window) with the best-fit stellar model.
The [\ion{O}{2}]-based and $L_u$-based SFR estimates are subject to different uncertainties.
As described in \S \ref{sec:application}, the [\ion{O}{2}]-based SFRs may be affected by object-to-object variation of the quasar contribution.
The dust extinction affects nebular lines and stellar continuum in different ways, due to the different geometry of the emission regions
\citep{calzetti94, charlot00}.
Nonetheless, Figure \ref{fig:sfrs} shows that the two estimates are roughly comparable, although there is a large scatter.
It may not be surprising that dust has only limited effect on these estimates. 
As previously mentioned, past observations demonstrated that the lines of sight toward optically-selected broad-line quasar nuclei are almost dust free; 
the color excess is less than $E_{B-V} = 0.1$ mag in the vast majority of cases \citep{richards03, hopkins04, salvato09, matute12, krawczyk14}.
A galaxy-wide dust screen is not likely to be present, although there could be localized dusty star-forming regions.
However, the present results indicate that the host galaxies are dominated by stars with intermediate ages and are not active star formers
(at least within the spectroscopic fibers).
Specific SFR (SFR/$M_*$) is even less sensitive to dust extinction than is SFR.
Our test indicates that assuming moderate extinction in the decomposition procedure (Models 5 and 6)
results in the stellar mass estimates simply increasing by the corresponding extinction factors.
This is because the extinction does not alter the derived stellar age and hence mass-to-luminosity ratio significantly.
As a result, extinction factors in SFR and $M_*$ are effectively cancelled out when specific SFR is derived.\footnote{
Little effect of dust extinction on the derived stellar age is expected from the narrow spectral window used in the first step of the decomposition procedure
($\lambda = 3700 - 4050$ \AA; see \S \ref{sec:methods}), in which the extinction cannot alter the spectral {\it slope} significantly.
However, the present argument may be complicated by the fact that nebular lines and stellar continuum are affected by the extinction in different ways, 
as described above.
}

We emphasize that our sample is not volume limited.
However, it includes most of the SDSS-RM quasars at $z < 1$, which are expected to be fairly complete down to the limiting magnitude (see \S 2).
Most of the initial sample is successfully processed through the decomposition analysis; the fraction of objects whose results we deem unreliable
($f_* < 0.1$ or $\Delta$ log $M_* > 0.5$ dex) is only 18 \%.
We provide the decomposition results for all the 191 quasars in the electronic edition of the journal.

\section{Discussion \label{sec:discussion}}


\subsection{Host galaxies of unobscured quasars \label{sec:hosts}}

\begin{figure*}
\epsscale{.9}
\plotone{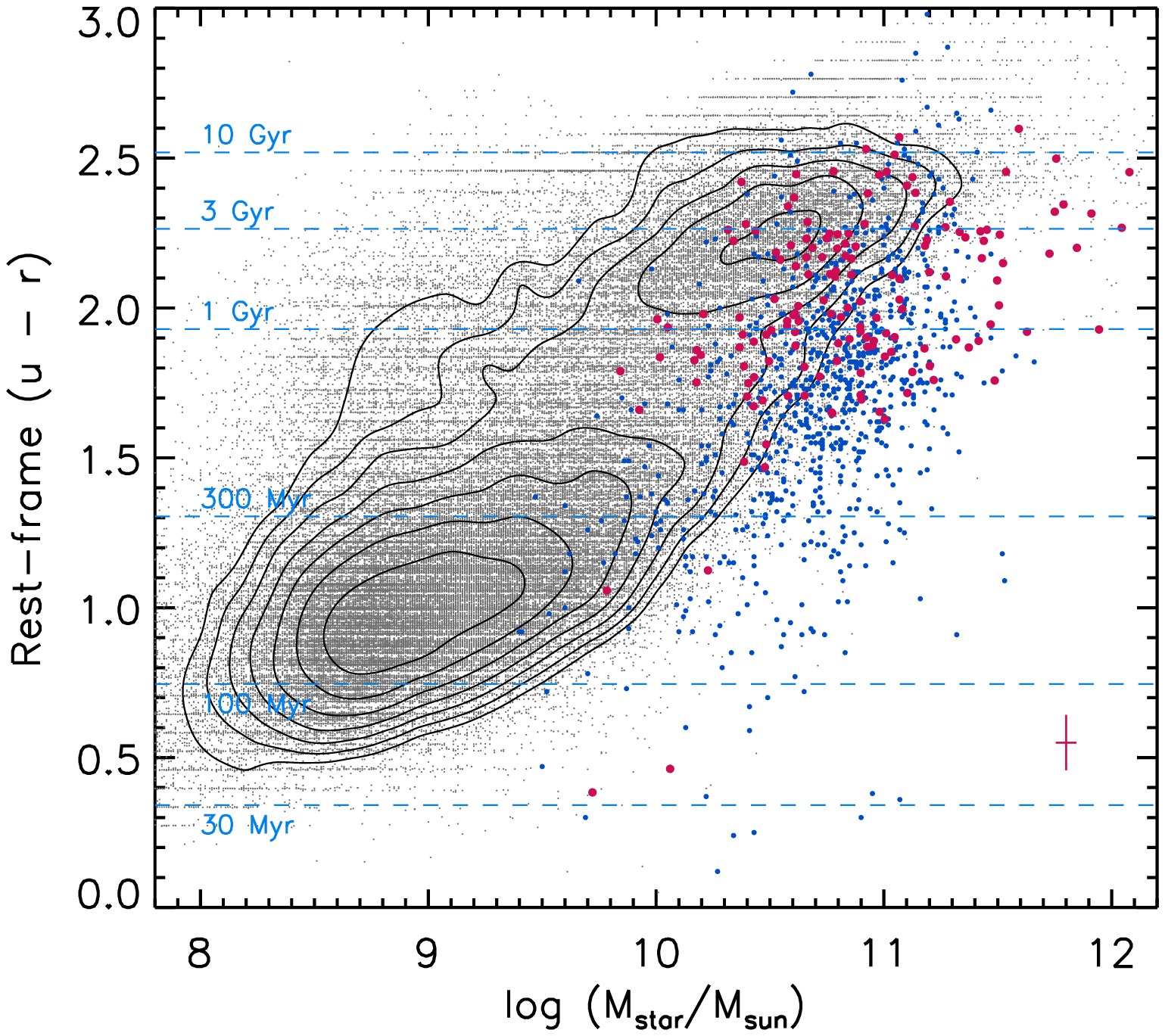}
\caption{Rest-frame ($u - r$) colors and stellar masses of the quasar host galaxies in this work (red dots) and in \citet[][blue dots]{matsuoka14}.
The typical error for the present sample is shown by the error bar at the bottom right corner.
The small gray dots and contours, drawn at logarithmically stepped levels of number density, represent non-AGN galaxies at $0.5 < z < 1.0$ 
taken from the COSMOS/UltraVISTA $K$-band selected catalog \citep{muzzin13}.
The dots are given small random offsets to improve visibility.
The dashed lines mark the SSP colors with $t_* = $ 0.03, 0.1, 0.3, 1, 3, and 10 Gyr, as labelled.
The quasar hosts are preferentially hosted by massive galaxies distributed from the massive tip of the blue cloud to the red sequence.
Their mean stellar age ($t_* \sim 1$ Gyr) corresponds to the gap of the bimodal distribution of inactive galaxies, where blue star-forming galaxies
may be rapidly transitioning to the red sequence.
\label{fig:cmd}
}
\end{figure*}

Before discussing the results, we first compare our measurements with those of \citet{matsuoka14}.
They studied host galaxies of SDSS broad-line quasars at $z < 0.6$ with the imaging decomposition technique (see \S 1), which uses completely different
data and methods from those in the present analysis.
The two studies share no common objects.
We calculated the host colors of the present sample by convolving the decomposed galaxy spectra (in the rest frame; they are extrapolated to $\lambda < 3700$ \AA,
i.e., outside our fitting window, with the best-fit stellar models) with the SDSS filter transmissions 
and created the CMD as shown in Figure \ref{fig:cmd}.
As a reference, we also plot the distribution of inactive galaxies at $0.5 < z < 1.0$ taken from the COSMOS/UltraVISTA $K$-band selected catalog \citep{muzzin13}; 
the rest-frame $u - r$ colors were calculated with the best-fit spectral model of each object given in the catalog. 
The two measurements of the quasar hosts broadly agree in distributions of stellar mass and color,
demonstrating that the quasars are preferentially hosted by massive galaxies distributed from the massive tip of the blue cloud to the red sequence.
The median stellar age of the present sample, $t_* \sim 1$ Gyr, corresponds to the gap of the bimodal color distribution of inactive galaxies.
It is intriguing that the quasars occupy the colors centered on this gap, 
where star-forming galaxies may be rapidly transitioning to the quiescent phase due to the quenching of star formation.
The stellar masses of the quasar hosts ($M_* \sim 10^{11} M_\odot$) broadly agree with the turn-over mass of the local
galaxy mass function at the high mass end \citep[e.g., ][]{bell03}.
Galaxy growth is significantly suppressed above this mass \citep{peng10}, for which quasar activity may be playing an important role.

\begin{figure}
\epsscale{1.15}
\plotone{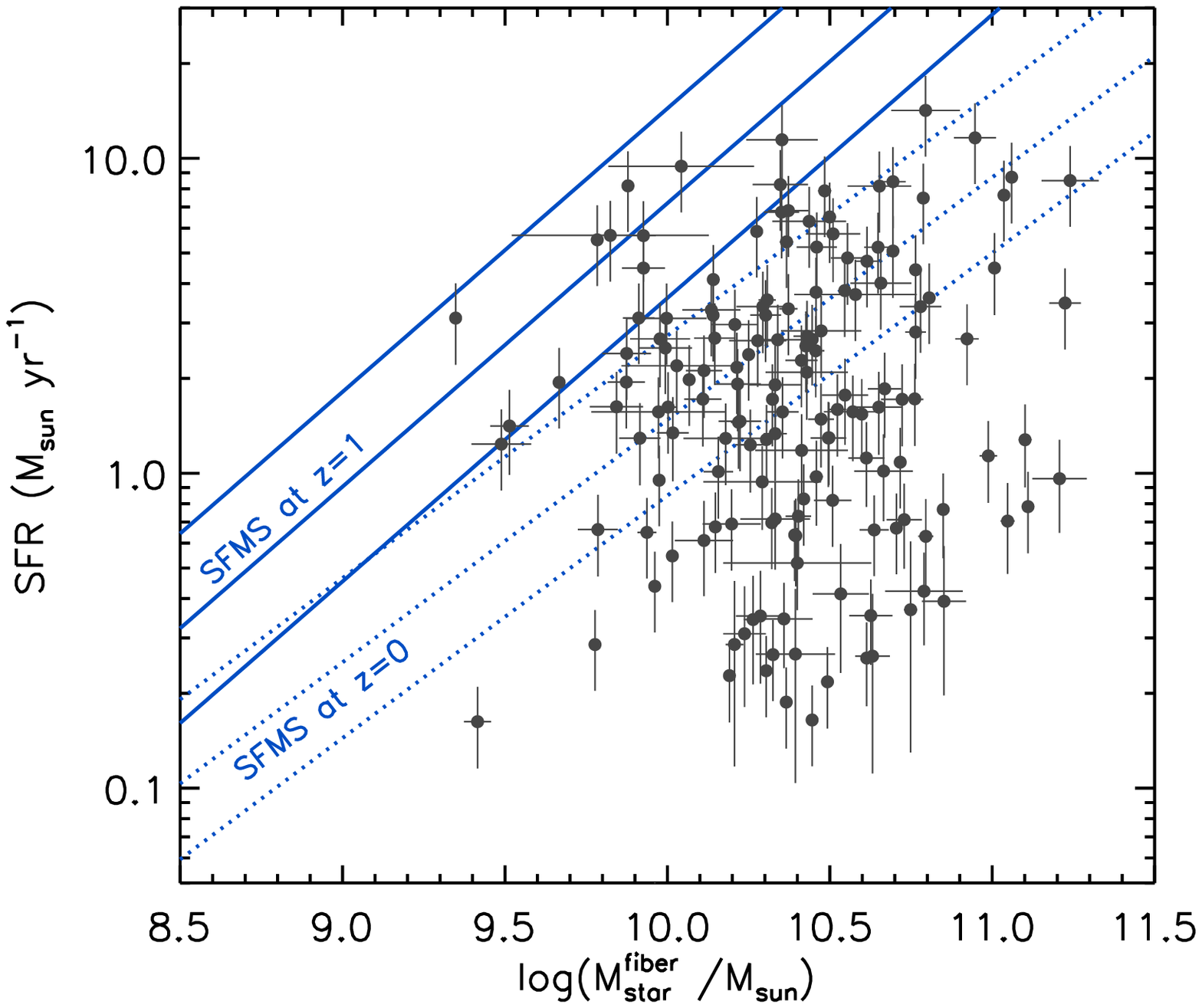}
\caption{Relation between SFRs and stellar masses within the 2\arcsec-diameter spectroscopic fibers. 
All the successful measurements of the 156 quasars are plotted.
The solid and dotted lines represent the main sequence of star-forming galaxies and their 68\% confidence intervals at $z = 1$ and $z = 0$, respectively,
taken from \citet{elbaz07}.
The quasar hosts fall below the main sequence, i.e., their star formation efficiencies are considerably lower than those of normal star-forming galaxies
at similar redshifts.
\label{fig:sfms}
}
\end{figure}

Figure \ref{fig:sfms} displays the relation between SFRs and stellar masses within the spectroscopic fibers,\footnote{
Here we use the stellar masses within the fibers (i.e., before the aperture correction) so that both stellar masses and SFRs refer to the same 
central region of the galaxies; see \S \ref{sec:application}.
} along with the so-called
``main sequence" of star-forming galaxies \citep[e.g., ][]{brinchmann04, noeske07, daddi07, elbaz07, elbaz11, steinhardt14}. 
The ratio between these two quantities defines specific SFR, which is not very sensitive to the dust extinction in the present analysis as discussed in \S 4.
The majority of the quasar hosts fall below the main sequence at $z = 1$, i.e., their star formation efficiencies are considerably lower than those of
normal star-forming galaxies at similar redshifts.
On the other hand, SFRs measured in host galaxies of dust-obscured, X-ray selected AGNs at similar redshifts tend to be higher than estimated 
here and trace the main sequence of star-forming galaxies \citep[][]{mullaney12, santini12}.
However, \citet{mullaney15} demonstrated recently that these SFR estimates are biased toward higher values due to bright outliers in stacking analysis of infrared data,
and that the actual SFR distribution of obscured AGNs is broader than, and has $\sim$0.4 dex lower peak of, that of main-sequence galaxies.
These findings are broadly consistent with our results for broad-line quasars presented in Figure \ref{fig:sfms}.

The intermediate stellar ages and relatively low SFRs of the present sample suggest that the quasar hosts are linked to post-starburst populations.
Since the first discovery by \citet{dressler83}, post-starburst galaxies have been recognized as an important stage of galaxy evolution.
They are characterized by the presence of strong Balmer absorption lines and the absence of emission lines, which are commonly
interpreted as evidence for starburst activity within the past $\sim$1 Gyr, which was quenched abruptly \citep[e.g., ][]{quintero04,goto05}.
The cause of this quenching is still unknown, but the present study indicates that the formation of at least some of post-starburst galaxies is related
to the nuclear activity. 
Quasars detected in post-starburst galaxies \citep[e.g.,][]{brotherton99,brotherton02} may represent the relevant phase.
In the \citet{hopkins06} merger-driven co-evolution model, a galaxy experiences various transition phases after a gas-rich merger, including a luminous infrared 
galaxy phase with intense star formation and dust production, an obscured AGN/quasar phase with activated central SMBHs, and an 
unobscured AGN/quasar phase before settling into a quiescent early-type galaxy.
Our results point to the link between quasar hosts and post-starburst galaxies, suggesting 
that merger-induced starbursts are largely quenched before the dust obscuring the nuclear region is cleared away and the central quasar becomes 
optically visible.
Recently \citet{pete15} estimated the number density of post-starburst galaxies to be roughly $10^{-6}$ Mpc$^{-3}$ mag$^{-1}$ for $M_{5000} > -20$ mag 
at $0 < z \la 1$, where $M_{5000}$ is the absolute magnitude at rest-frame 5000 \AA.
This value is roughly comparable to the number density of broad-line quasars at similar redshifts and luminosities \citep{richards06,croom09}.
\citet{kauffmann03} noted that the strong H$\delta$ absorption observed in low-$z$ ($z \sim 0.05$) SDSS narrow-line AGNs indicates
they experienced a burst of star formation that ended within the past $\sim$1 Gyr.
\citet{yesuf14} demonstrated that the AGN fraction is three times higher in post-starburst galaxies than in inactive galaxies.
However, they also reported that (obscured) AGN phase is delayed by a few 100 Myr relative to post-starburst phase, and hence cannot play 
a primary role in the quenching of starbursts.

\begin{figure}
\epsscale{1.1}
\plotone{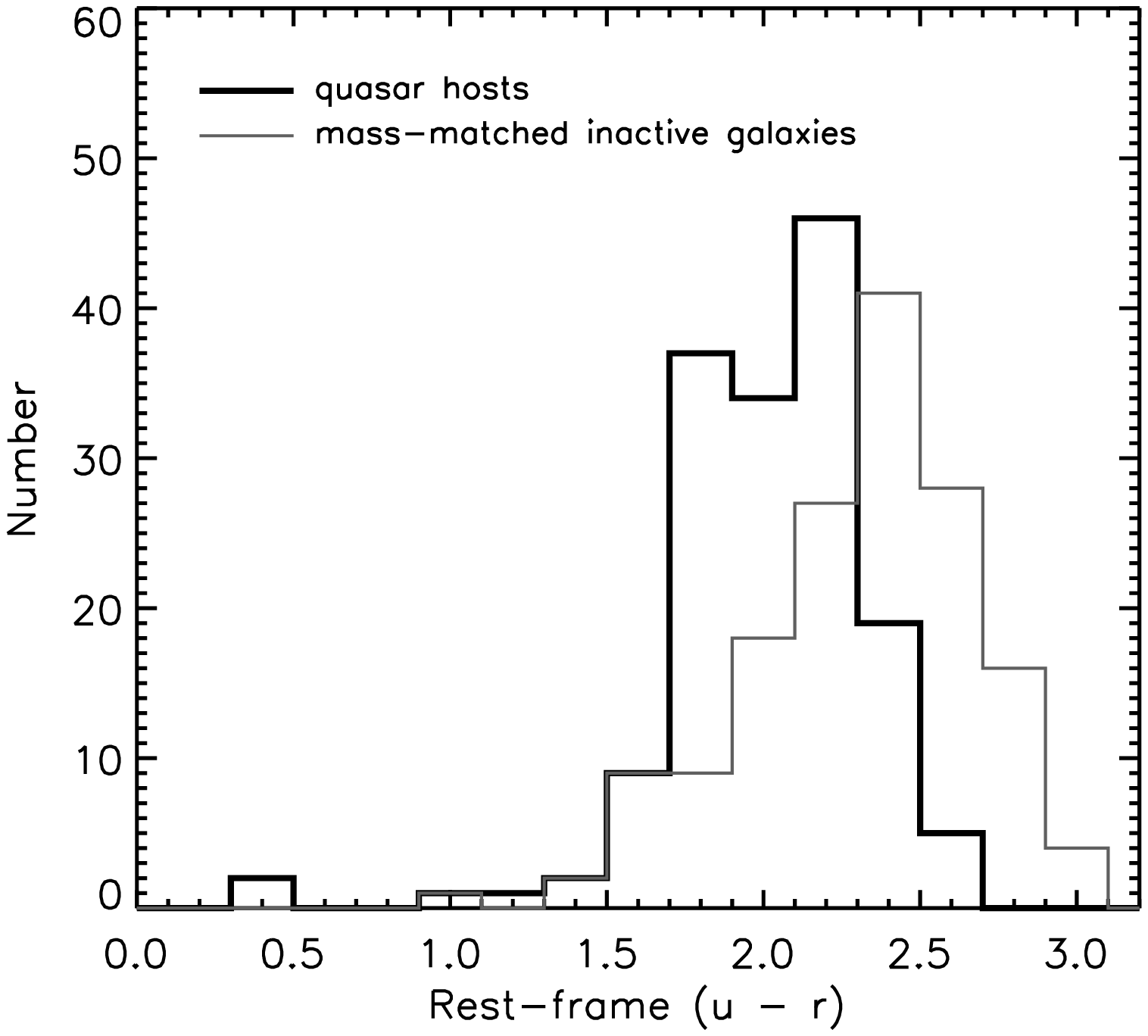}
\caption{Histograms of the rest-frame ($u - r$) colors of the quasar host galaxies in this work (thick line) and the mass-matched inactive (non-AGN) galaxies 
drawn from the COSMOS/UltraVISTA catalog (thin line).
The quasar host distribution is shifted toward bluer colors than that of inactive galaxies, demonstrating that bluer galaxies are more likely to host quasars.
\label{fig:agnfrac}
}
\end{figure}

There have been a number of studies comparing the color and mass distributions of AGN host galaxies with those of inactive galaxies 
\citep[e.g.,][]{nandra07, salim07,hickox09, schawinski10}.
Recent work suggests that, using proper mass-limited samples, the AGN fraction is constant as a function of  host color, or may be enhanced 
in blue star-forming galaxies \citep[e.g.,][]{silverman09, aird12, hainline12, rosario13}.
Luminous AGNs are preferentially found in massive host galaxies, which is perhaps due to the presence of massive SMBHs at the centers and/or
of large reservoirs of gas in such galaxies.
Most of these studies are based on obscured AGNs selected via their X-ray luminosity or optical narrow emission lines.
Our work (Figure \ref{fig:cmd}) suggests that broad-line quasar hosts prevail in the color range $1.5 < u - r < 2.5$ above masses of 
log $(M_*/M_\odot) \sim 10.5$ and that the quasar fraction increases toward bluer colors, a trend that continues to the end of the galaxy color distribution
at $u - r < 1.5$.
In order to show this trend more clearly, we create a mass-matched sample of inactive galaxies; for each of the quasar hosts, the galaxy with
the closest mass and redshift are drawn from the COSMOS/UltraVISTA $K$-band selected catalog.
Figure \ref{fig:agnfrac} compares the histograms of the rest-frame ($u - r$) colors of the two samples. 
The quasar host distribution is shifted toward bluer colors than that of mass-matched inactive galaxies, demonstrating that bluer galaxies are more
likely to host quasars.

\begin{figure*}
\epsscale{1.1}
\plotone{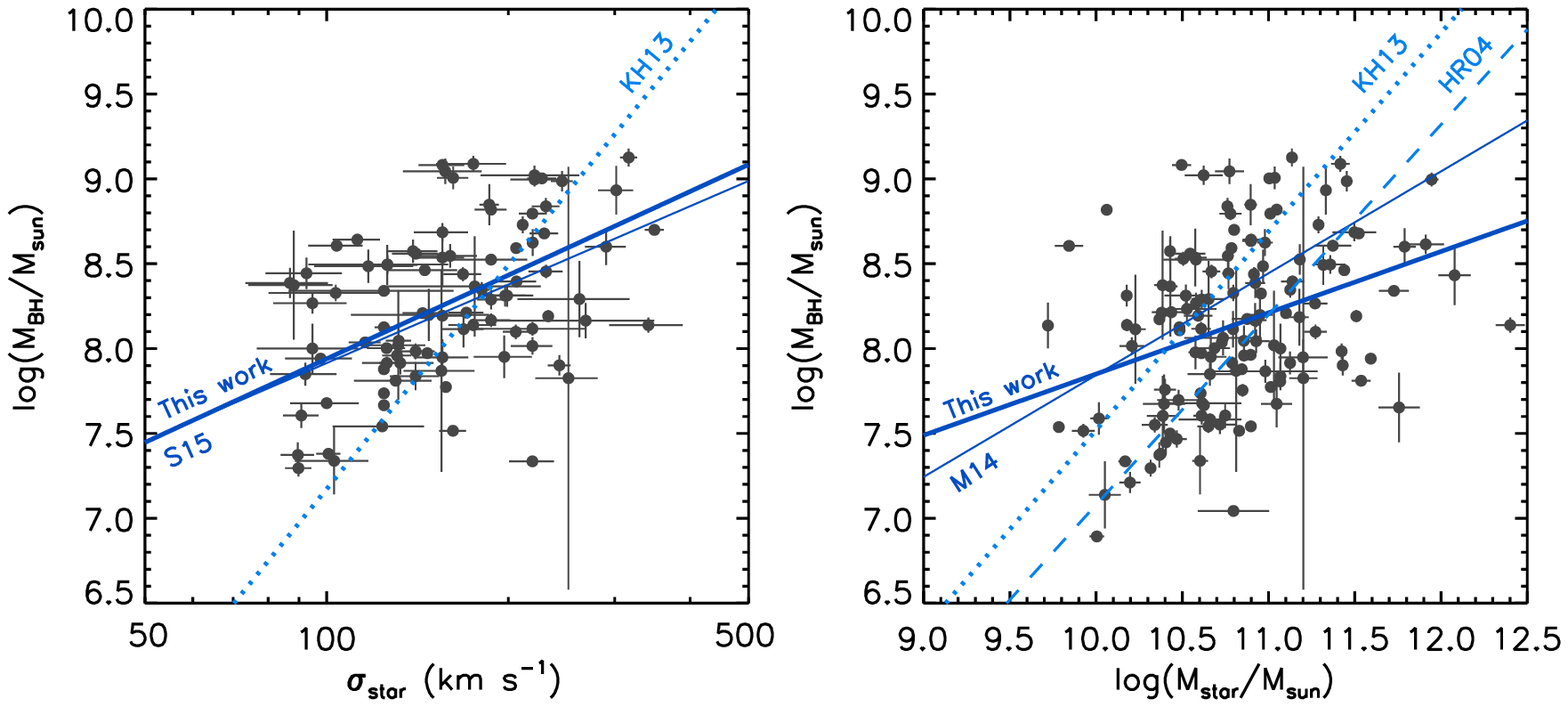}
\caption{Relation between $M_\bullet$ and $\sigma_*$ (left) and between $M_\bullet$ and $M_*$ (right).
The thick solid lines represent the regression lines of the present sample, while the thin solid lines indicate those of the previous measurements 
by \citet[][``S15"]{shen15} and \citet[][``M14"]{matsuoka14, matsuoka14e}.
The dotted and dashed lines represent the local relations taken from \citet[][``KH13"]{kormendy13} and \citet[][``HR04"]{haring04}, respectively.
Note that the KH13 and HR04 relations have been calibrated with bulge masses, while $M_*$ of this work includes a galaxy disk component if it is present (see text).
The mean $M_\bullet$/$\sigma_*$ or $M_\bullet$/$M_*$ ratio of the quasar hosts is consistent with the local value, while the slope of the relation
appears to flatten toward high redshift.
This flattening can be explained by selection biases (see text).
\label{fig:msigma}
}
\end{figure*}

Meanwhile, a drawback of the present analysis is the lack of far-infrared information and hence the dust-absorbed part of star formation.
Although a galaxy-wide dust screen is unlikely to be present (see \S 4), the present work cannot trace localized star-forming regions 
hidden by dust at UV-to-optical wavelengths.
Such regions can be mapped reliably only with far-infrared observations, which we will exploit in future work.
Our analysis is also limited to within the spectroscopic fibers, whose 2\arcsec\ diameter corresponds to approximately 14 kpc at the median redshift 
of the sample (see \S \ref{sec:application}).
If the majority of the star formation took place at outside this aperture, then our results would become inconsistent with the past studies which measured
 the entire host galaxies.


\subsection{Scaling relations}

Our analysis provides both stellar velocity dispersions and masses as well as SMBH masses, 
which allows us to investigate the scaling relations between them.
Figure \ref{fig:msigma} presents the $M_\bullet - \sigma_*$ and $M_\bullet - M_*$ relations of the present sample.
We calculate the linear regression lines with the IDL routine \texttt{MPFITEXY}, 
as:
\begin{eqnarray}
	{\rm log} \biggl( \frac{M_\bullet}{10^8 M_\odot} \biggr) & = & (1.64 \pm 0.29)\ {\rm log}  \biggl( \frac{\sigma_*}{200\ {\rm km\ s}^{-1}}  \biggr) \nonumber \\
		& & + (0.43 \pm 0.05) ,
\end{eqnarray}
and
\begin{eqnarray}
	{\rm log} \biggl( \frac{M_\bullet}{10^8 M_\odot} \biggr) & = &  (0.36 \pm 0.08)\ {\rm log}  \biggl( \frac{M_*}{10^{11} M_\odot}  \biggr) \nonumber \\
                  & & + (0.21 \pm 0.04) .
\end{eqnarray}
The intrinsic scatters are 0.39 and 0.44 dex for the $M_\bullet - \sigma_*$ and $M_\bullet - M_*$ relations, respectively.
The $M_\bullet - \sigma_*$ relation is almost identical to that of \citet[][see \S \ref{sec:simulation}]{shen15}, which is not surprising
given the consistent measurements of the two quantities (Figure \ref{fig:shen15}).
Our sample shows no evolution of the mean $M_\bullet$/$\sigma_*$ ratio from the local Universe, while the slope of the relation appears 
to flatten toward high redshifts.
This flattening was also reported by \citet{shen15} and was interpreted as a result of a selection bias; with simple simulations,
they concluded that a non-evolving $M_\bullet - \sigma_*$ relation is favored at $z < 1$.

We also found similar trends in the $M_\bullet - M_*$ relation.
Our $M_*$ estimates include a galaxy disk component if it is present, hence they should be regarded as the upper limits of the bulge masses ($M_{\rm *, bulge}$).
A similar comparison between the $M_\bullet - M_*$ relation at $z \sim 1.4$ and the local $M_\bullet - M_{\rm *, bulge}$ relation was presented 
by \citet{jahnke09}, who found no significant difference between the two.
Since their sample shows evidence for substantial disk components, the authors concluded that bulges have grown predominantly via redistribution of the disk mass 
since $z = 1.4$ to the present epoch.
This conclusion was supported by \citet{sun15} with a larger sample at similar redshifts and with a careful assessment of selection biases.
Our results are consistent with these findings, but we have no estimate of disk contribution to $M_*$ and cannot provide a direct support for the above scenario.
The present $M_\bullet - M_*$ relation is also similar to that of \citet{matsuoka14}, who reported that the relation is offset toward large $M_\bullet$
compared to the local relation at fixed $M_*$.
This result is likely due to the shallow slope of the relation, producing systematically large $M_\bullet$/$M_*$ ratios at $M_* \la 10^{11} M_\odot$. 

\subsection{Applicability of the present analysis}

The present spectral decomposition analysis is made possible due to the low-to-moderate luminosities of the quasars,
with relative stellar contribution $f_* > 0.1$ measured at $\lambda = 4000$ \AA, and the high SNR of the spectra.
Meeting both conditions usually requires significant telescope resources; in this work, we have exploited the co-added SDSS-RM spectra whose net
exposure times amount to 65 hours per object.
As we demonstrated in the previous sections, clear detection of stellar absorption features allows for reliable extraction of 
the host components from the observed spectra.
In this section, we give a rough guideline for estimating whether the present analysis is applicable to a given spectrum.

\begin{figure}
\epsscale{1.1}
\plotone{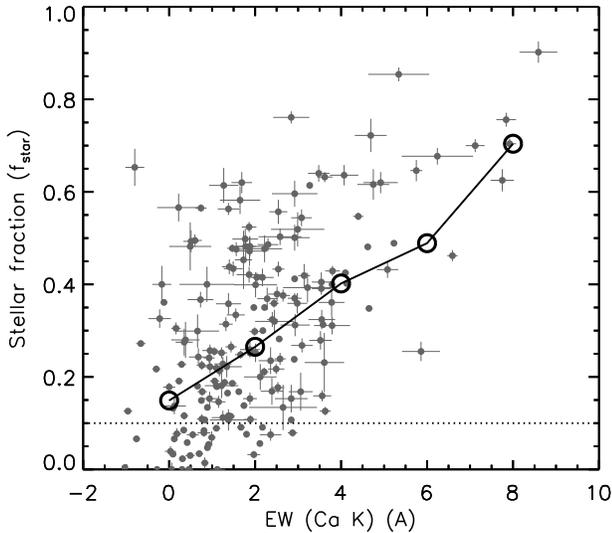}
\caption{
Stellar fraction ($f_*$) measured at $\lambda$ = 4000 \AA\ in our sample of 191 quasars, as a function of the rest-frame EW (Ca K); a positive EW value means absorption.
The large circles represent median values in bins of EW (Ca K).
The dotted line marks the success threshold of our decomposition analysis, $f_* > 0.1$.
There is a moderate correlation between the two quantities, which allows one to make a rough estimate of $f_*$ by measuring EW (Ca K) in a given spectrum.
\label{fig:CaKFssp}
}
\end{figure}

The first requirement is a certain fraction ($f_* \ga 0.1$) of host galaxy light in the total spectrum.
This fraction can be estimated roughly from the depths of the stellar absorption lines.
Figure \ref{fig:CaKFssp} presents the stellar fraction $f_*$ 
as a function of EW (Ca K), the rest-frame equivalent width of the Ca K absorption line.
Ca K is one of the strongest stellar absorption features in a quasar spectrum at UV-to-optical wavelengths, while Ca H
is less visible due to the superposed [\ion{Ne}{3}] $\lambda$3969 emission. 
We measure EW (Ca K) in the wavelength range $\lambda$ = 3920 -- 3955 \AA, with the blue and red continuum levels defined 
at $\lambda$ = 3880 -- 3920 \AA\ and $\lambda$ = 3990 -- 4030 \AA, respectively.
Figure \ref{fig:CaKFssp} indicates that EW (Ca K) and $f_*$ are moderately correlated.
The majority of the quasars with EW (Ca K) $\ga$ 2 \AA\ have more than 10\% of the light at 4000 \AA\ from the host, and 
stellar emission is the major contributor to a quasar spectrum on average when one observes EW (Ca K) $\ga$ 6 \AA.

\begin{figure*}
\epsscale{.9}
\plotone{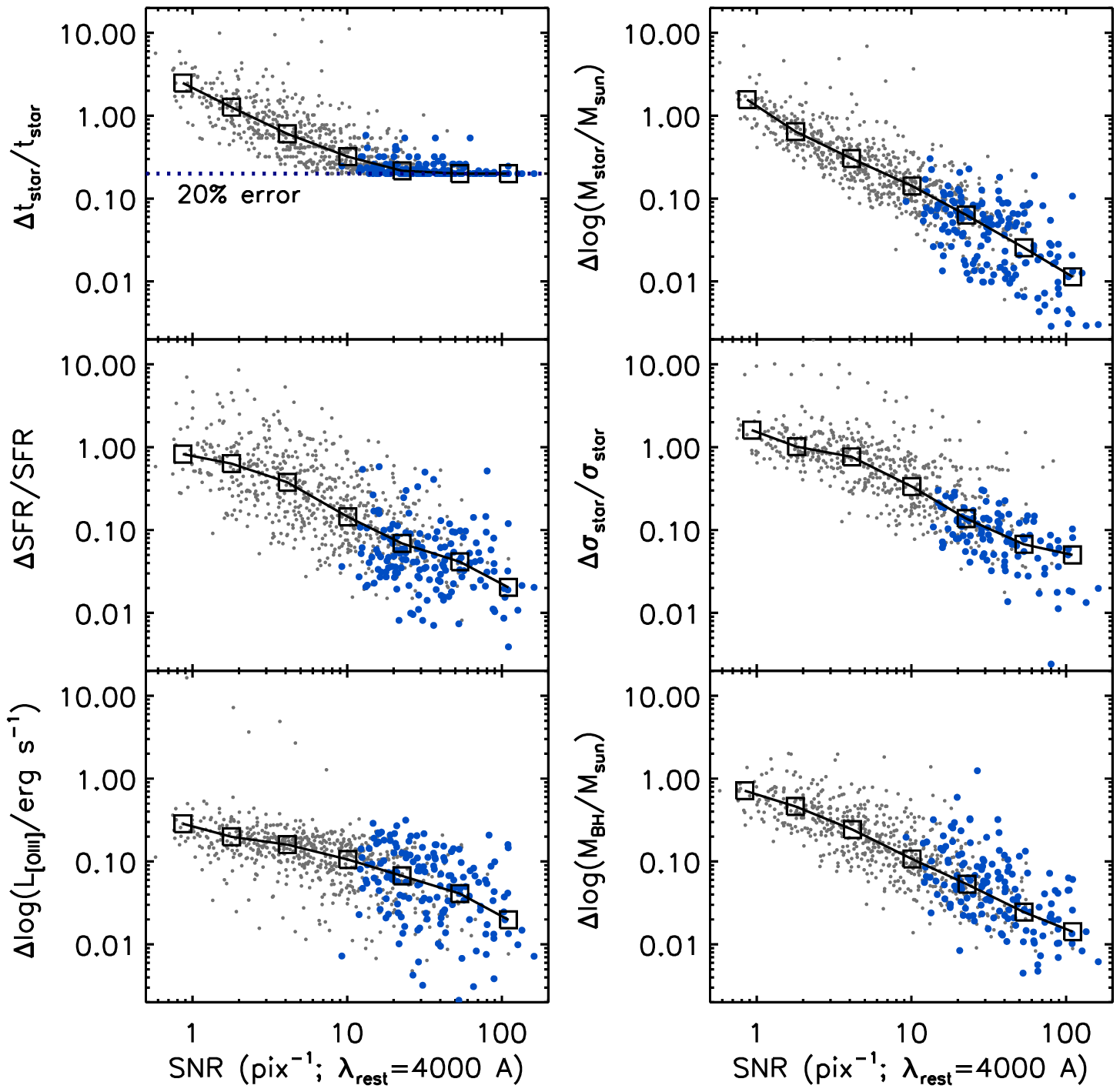}
\caption{
Uncertainties in $t_*$ (top left), $M_*$ (top right), SFR, (middle left), $\sigma_*$ (middle right), $L_{\rm [OIII]}$ (bottom left), and $M_\bullet$ (bottom right),
as a function of the SNR per pixel measured at $\lambda$ = 4000 \AA.
The large dots represent the original SDSS-RM spectra, while the small dots represent the mock spectra with noise added.
The quantity $\Delta t_*/t_*$ has a lower limit at 0.2 (dotted line in the upper left panel) because of the 20 \% error added to account for non-contiguous
age coverage of the SSP models (see \S \ref{sec:application}).
The squares connected by the solid lines show the median values in bins of SNR.
\label{fig:snred}
}
\end{figure*}

The second requirement is high SNR.
We test the minimum SNR required to perform a reliable decomposition analysis 
by artificially degrading the SDSS-RM data and determining how the derived uncertainties change as the SNR decreases.
We take the original high-SNR spectra with successful decomposition and add noise assuming a Gaussian PDF with standard 
deviations of 2, 4, 8, or 16 times the original pixel errors.
These mock spectra are processed through the decomposition algorithm. 
We omit the calibration errors in Equations \ref{eq:sfr} and \ref{eq:mbh} in these simulations to view the effect of decreasing SNR directly.
Figure \ref{fig:snred} displays the resultant uncertainties in the six physical quantities as a function of the SNR per pixel measured at $\lambda$ = 4000 \AA.
These plots can be used to infer the required SNR to achieve the desired accuracy in one of the parameters;
for example, a rough estimate of stellar mass ($\Delta$log $M_* < 0.5$ dex) is possible at SNR $\sim 3$ on average, while stellar age is poorly constrained
($\Delta t_*/t_* \sim 1$) at that SNR.
All six parameters are moderately well constrained at SNR $>$ 10, where the actual SDSS-RM quasars are located.

\section{Summary \label{sec:summary}}

We present the results of a spectral decomposition analysis of broad-line quasars at $z < 1$.
The sample is drawn from the SDSS-RM project, in which a single spectroscopic field was repeatedly observed with the BOSS spectrograph to explore the variability of quasars.
The high SNR co-added spectra of 191 SDSS-RM quasars, whose net exposure times amount to approximately 65 hours per object, are used for the analysis.
We develop a technique to decompose a quasar spectrum into nuclear and host components using spectral models of the quasar accretion disk,
gas in the BLR, NLR, ISM, and stars in the host galaxy.
This procedure provides estimates of stellar age and mass, SFR, stellar velocity dispersion, quasar [\ion{O}{3}] luminosity, and SMBH mass, among other quantities.
The decomposition was successful for more than 80 \% of the initial sample; 
these objects have more than 10 \% of their light at 4000 \AA\ from the host, and the error in stellar mass estimates $\Delta$log $(M_*/M_\odot)$ is less than 0.5.
We find that the quasars are preferentially hosted by massive ($M_* \sim 10^{11} M_\odot$) galaxies with characteristic stellar ages around $t_*$ = 1 Gyr.
It coincides with the gap of the bimodal color distribution of inactive galaxies, 
where blue star-forming galaxies may be rapidly transitioning into the red sequence.
The deficit of young stars is consistent with the estimated low SFRs, which place the quasar hosts below the main sequence of star-forming galaxies at similar redshifts.
These facts suggest that the quasar hosts have experienced an episode of major star formation sometime in the past $\sim$1 Gyr, which was subsequently
quenched or suppressed (at a slower rate) .
We also find that the scaling relations between $M_\bullet$  and $\sigma_*$ and between $M_\bullet$ and $M_*$ are consistent with our previous measurements.
The mean $M_\bullet$/$\sigma_*$ or $M_\bullet/M_*$ ratio is consistent with the local value, while the slope of the relation is flatter than the local relation.
As demonstrated by \citet{shen15}, this result is likely due to a selection bias, and no evolution of the scaling relations is favored at $z < 1$.
Finally, we demonstrate that the stellar contribution to a quasar spectrum can be inferred roughly from the equivalent widths of Ca K absorption line, and 
provide estimates of the minimum SNR required to perform a spectral decomposition with desired precision. 

The present work will be greatly advanced with future powerful spectrographs such as the Subaru Prime Focus Spectrograph \citep[PFS; ][]{takada14},
which will produce large numbers of high-quality spectra of galaxies and quasars out to high redshifts.
Our method, optimized for BOSS fiber spectra, can be readily applied to other high-quality spectra in order to study quasars 
with lower luminosity and higher redshifts.
It is also important to extend the present analysis into longer rest-frame wavelengths ($\lambda > 5400$ \AA) with near-infrared spectrographs, 
where the relative host contribution becomes larger and the possible effects of dust extinction become smaller.

\acknowledgments

We are grateful to the referee for his/her useful comments and suggestions.
YM thanks Nobunari Kashikawa for fruitful discussions.
WNB thanks NSF grant AST-1108604 for support.
LCH acknowledges support by the Chinese Academy of Science through grant No. XDB09030102 (Emergence of Cosmological Structures) from the Strategic Priority Research Program and by the National Natural Science Foundation of China through grant No. 11473002.
Support for the work of YS was provided by NASA through Hubble Fellowship grant number HST-HF-51314, awarded by the Space Telescope Science Institute, 
which is operated by the Association of Universities for Research in Astronomy, Inc., for NASA, under contract NAS 5-26555.
MYS acknowledges support from the China Scholarship Council (No. [2013]3009).


Funding for SDSS-III has been provided by the Alfred P. Sloan Foundation, the Participating Institutions, the National Science Foundation, and the U.S. Department of Energy Office of Science. The SDSS-III Web site is http://www.sdss3.org/.

SDSS-III is managed by the Astrophysical Research Consortium for the Participating Institutions of the SDSS-III Collaboration including the University of Arizona, the Brazilian Participation Group, Brookhaven National Laboratory, University of Cambridge, Carnegie Mellon University, University of Florida, the French Participation Group, the German Participation Group, Harvard University, the Instituto de Astrofisica de Canarias, the Michigan State/Notre Dame/JINA Participation Group, Johns Hopkins University, Lawrence Berkeley National Laboratory, Max Planck Institute for Astrophysics, Max Planck Institute for Extraterrestrial Physics, New Mexico State University, New York University, Ohio State University, Pennsylvania State University, University of Portsmouth, Princeton University, the Spanish Participation Group, University of Tokyo, University of Utah, Vanderbilt University, University of Virginia, University of Washington, and Yale University.


\begin{thebibliography}{}
\bibitem[Abazajian et al.(2004)]{abazajian04} Abazajian, K., Adelman-McCarthy, J.~K., Ag{\"u}eros, M.~A., et al.\ 2004, \aj, 128, 502 
\bibitem[Aird et al.(2012)]{aird12} Aird, J., Coil, A.~L., Moustakas, J., et al.\ 2012, \apj, 746, 90 
\bibitem[Alam et al.(2015)]{alam15} Alam, S., Albareti, F.~D., Allende Prieto, C., et al.\ 2015, arXiv:1501.00963 
\bibitem[Annis et al.(2014)]{annis14} Annis, J., Soares-Santos, M., Strauss, M.~A., et al.\ 2014, \apj, 794, 120 
\bibitem[Bahcall et al.(1997)]{bahcall97} Bahcall, J.~N., Kirhakos, S., Saxe, D.~H., \& Schneider, D.~P.\ 1997, \apj, 479, 642 
\bibitem[Ba{\~n}ados et al.(2014)]{banados14} Ba{\~n}ados, E., Venemans, B.~P., Morganson, E., et al.\ 2014, \aj, 148, 14 
\bibitem[Barnes \& Hernquist(1991)]{barnes91} Barnes, J.~E., \& Hernquist, L.~E.\ 1991, \apjl, 370, L65 
\bibitem[Bell et al.(2003)]{bell03} Bell, E.~F., McIntosh, D.~H., Katz, N., \& Weinberg, M.~D.\ 2003, \apjs, 149, 289 
\bibitem[Bernardi et al.(2003)]{bernardi03} Bernardi, M., Sheth, R.~K., Annis, J., et al.\ 2003, \aj, 125, 1817 
\bibitem[Bolton et al.(2012)]{bolton12} Bolton, A.~S., Schlegel, D.~J., Aubourg, {\'E}., et al.\ 2012, \aj, 144, 144 
\bibitem[Boroson \& Oke(1982)]{boroson82} Boroson, T.~A., \& Oke, J.~B.\ 1982, \nat, 296, 397 
\bibitem[Bower et al.(2006)]{bower06} Bower, R.~G., Benson, A.~J., Malbon, R., et al.\ 2006, \mnras, 370, 645 
\bibitem[Brandt \& Alexander(2015)]{brandt15} Brandt, W.~N., \& Alexander, D.~M.\ 2015, \aapr, 23, 1 
\bibitem[Brinchmann et al.(2004)]{brinchmann04} Brinchmann, J., Charlot, S., White, S.~D.~M., et al.\ 2004, \mnras, 351, 1151 
\bibitem[Brotherton et al.(2002)]{brotherton02} Brotherton, M.~S., Grabelsky, M., Canalizo, G., et al.\ 2002, \pasp, 114, 593 
\bibitem[Brotherton et al.(1999)]{brotherton99} Brotherton, M.~S., van Breugel, W., Stanford, S.~A., et al.\ 1999, \apjl, 520, L87 
\bibitem[Calzetti et al.(1994)]{calzetti94} Calzetti, D., Kinney, A.~L., \& Storchi-Bergmann, T.\ 1994, \apj, 429, 582 
\bibitem[Canalizo \& Stockton(2013)]{canalizo13} Canalizo, G., \& Stockton, A.\ 2013, \apj, 772, 132 
\bibitem[Cano-D{\'{\i}}az et al.(2012)]{cano-diaz12} Cano-D{\'{\i}}az, M., Maiolino, R., Marconi, A., et al.\ 2012, \aap, 537, L8 
\bibitem[Cappellari \& Emsellem(2004)]{cappellari04} Cappellari, M., \& Emsellem, E.\ 2004, \pasp, 116, 138 
\bibitem[Cardamone et al.(2010)]{cardamone10} Cardamone, C.~N., Urry, C.~M., Schawinski, K., et al.\ 2010, \apjl, 721, L38 
\bibitem[Chabrier(2003)]{chabrier03} Chabrier, G.\ 2003, \pasp, 115, 763 
\bibitem[Charlot \& Fall(2000)]{charlot00} Charlot, S., \& Fall, S.~M.\ 2000, \apj, 539, 718 
\bibitem[Choi \& Nagamine(2011)]{choi11} Choi, J.-H., \& Nagamine, K.\ 2011, \mnras, 410, 2579 
\bibitem[Comastri et al.(2015)]{comastri15} Comastri, A., Gilli, R., Marconi, A., Risaliti, G., \& Salvati, M.\ 2015, \aap, 574, L10 
\bibitem[Croom et al.(2009)]{croom09} Croom, S.~M., Richards, G.~T., Shanks, T., et al.\ 2009, \mnras, 399, 1755 
\bibitem[Croton et al.(2006)]{croton06} Croton, D.~J., Springel, V., White, S.~D.~M., et al.\ 2006, \mnras, 365, 11 
\bibitem[Daddi et al.(2007)]{daddi07} Daddi, E., Dickinson, M., Morrison, G., et al.\ 2007, \apj, 670, 156 
\bibitem[Dav{\'e} et al.(2011)]{dave11} Dav{\'e}, R., Oppenheimer, B.~D., \& Finlator, K.\ 2011, \mnras, 415, 11 
\bibitem[Dawson et al.(2013)]{dawson13} Dawson, K.~S., Schlegel, D.~J., Ahn, C.~P., et al.\ 2013, \aj, 145, 10 
\bibitem[Di Matteo et al.(2005)]{dimatteo05} Di Matteo, T., Springel, V., \& Hernquist, L.\ 2005, \nat, 433, 604 
\bibitem[DiPompeo et al.(2014)]{dipompeo14} DiPompeo, M.~A., Myers, A.~D., Hickox, R.~C., Geach, J.~E., \& Hainline, K.~N.\ 2014, \mnras, 442, 3443 
\bibitem[DiPompeo et al.(2015)]{dipompeo15} DiPompeo, M.~A., Myers, A.~D., Hickox, R.~C., et al.\ 2015, \mnras, 446, 3492 
\bibitem[Dressler(1989)]{dressler89} Dressler, A.\ 1989, Active Galactic Nuclei: proceedings of IAUS 134, held in Santa Cruz, California, August 15-19, 1988. Edited by Donald E. Osterbrock and Joseph S. Miller. IAUS 134, Kluwer Academic Publishers, Dordrecht, p.217
\bibitem[Dressler \& Gunn(1983)]{dressler83} Dressler, A., \& Gunn, J.~E.\ 1983, \apj, 270, 7 
\bibitem[Dunlop et al.(2003)]{dunlop03} Dunlop, J.~S., McLure, R.~J., Kukula, M.~J., et al.\ 2003, \mnras, 340, 1095 
\bibitem[Elbaz et al.(2007)]{elbaz07} Elbaz, D., Daddi, E., Le Borgne, D., et al.\ 2007, \aap, 468, 33 
\bibitem[Elbaz et al.(2011)]{elbaz11} Elbaz, D., Dickinson, M., Hwang, H.~S., et al.\ 2011, \aap, 533, AA119 
\bibitem[Elitzur(2012)]{elitzur12} Elitzur, M.\ 2012, \apjl, 747, L33 
\bibitem[Eisenstein et al.(2011)]{eisenstein11} Eisenstein, D.~J., Weinberg, D.~H., Agol, E., et al.\ 2011, \aj, 142, 72 
\bibitem[Fabian(2012)]{fabian12} Fabian, A.~C.\ 2012, \araa, 50, 455 
\bibitem[Fan et al.(2006)]{fan06} Fan, X., Strauss, M.~A., Becker, R.~H., et al.\ 2006, \aj, 132, 117 
\bibitem[Ferrara et al.(2014)]{ferrara14} Ferrara, A., Salvadori, S., Yue, B., \& Schleicher, D.\ 2014, \mnras, 443, 2410 
\bibitem[Ferrarese \& Merritt(2000)]{ferrarese00} Ferrarese, L., \& Merritt, D.\ 2000, \apjl, 539, L9 
\bibitem[Feruglio et al.(2010)]{feruglio10} Feruglio, C., Maiolino, R., Piconcelli, E., et al.\ 2010, \aap, 518, L155 
\bibitem[Francis(1996)]{francis96} Francis, P.~J.\ 1996, PASA, 13, 212 
\bibitem[Fu \& Stockton(2009)]{fu09} Fu, H., \& Stockton, A.\ 2009, \apj, 690, 953 
\bibitem[Fukugita et al.(1996)]{fukugita96} Fukugita, M., Ichikawa, T., Gunn, J.~E., et al.\ 1996, \aj, 111, 1748 
\bibitem[Ganguly et al.(2007)]{ganguly07} Ganguly, R., Brotherton, M.~S., Cales, S., et al.\ 2007, \apj, 665, 990 
\bibitem[Gebhardt et al.(2000)]{gebhardt00} Gebhardt, K., Bender, R., Bower, G., et al.\ 2000, \apjl, 539, L13 
\bibitem[Georgakakis et al.(2008)]{georgakakis08} Georgakakis, A., Nandra, K., Yan, R., et al.\ 2008, \mnras, 385, 2049 
\bibitem[Glikman et al.(2012)]{glikman12} Glikman, E., Urrutia, T., Lacy, M., et al.\ 2012, \apj, 757, 51 
\bibitem[Gofford et al.(2013)]{gofford13} Gofford, J., Reeves, J.~N., Tombesi, F., et al.\ 2013, \mnras, 430, 60 
\bibitem[Goto(2005)]{goto05} Goto, T.\ 2005, \mnras, 357, 937 
\bibitem[Greene \& Ho(2006)]{greene06} Greene, J.~E., \& Ho, L.~C.\ 2006, \apj, 641, 117 
\bibitem[Greene \& Ho(2005)]{greene05} Greene, J.~E., \& Ho, L.~C.\ 2005, \apj, 627, 721 
\bibitem[Greene et al.(2011)]{greene11} Greene, J.~E., Zakamska, N.~L., Ho, L.~C., \& Barth, A.~J.\ 2011, \apj, 732, 9 
\bibitem[G{\"u}ltekin et al.(2009)]{gultekin09} G{\"u}ltekin, K., Richstone, D.~O., Gebhardt, K., et al.\ 2009, \apj, 698, 198 
\bibitem[Gunn et al.(2006)]{gunn06} Gunn, J.~E., Siegmund, W.~A., Mannery, E.~J., et al.\ 2006, \aj, 131, 2332 
\bibitem[Hainline et al.(2012)]{hainline12} Hainline, K.~N., Shapley, A.~E., Greene, J.~E., et al.\ 2012, \apj, 760, 74 
\bibitem[Hao et al.(2005)]{hao05} Hao, L., Strauss, M.~A., Fan, X., et al.\ 2005, \aj, 129, 1795 
\bibitem[H{\"a}ring \& Rix(2004)]{haring04} H{\"a}ring, N., \& Rix, H.-W.\ 2004, \apjl, 604, L89 
\bibitem[Hewett \& Wild(2010)]{hewett10} Hewett, P.~C., \& Wild, V.\ 2010, \mnras, 405, 2302 
\bibitem[Hickox et al.(2009)]{hickox09} Hickox, R.~C., Jones, C., Forman, W.~R., et al.\ 2009, \apj, 696, 891 
\bibitem[Ho(2009)]{ho09} Ho, L.~C.\ 2009, \apj, 699, 638 
\bibitem[Ho(2005)]{ho05} Ho, L.~C.\ 2005, \apj, 629, 680 
\bibitem[Hopkins et al.(2003)]{hopkins03} Hopkins, A.~M., Miller, C.~J., Nichol, R.~C., et al.\ 2003, \apj, 599, 971 
\bibitem[Hopkins et al.(2006)]{hopkins06} Hopkins, P.~F., Hernquist, L., Cox, T.~J., et al.\ 2006, \apjs, 163, 1 
\bibitem[Hopkins et al.(2007)]{hopkins07} Hopkins, P.~F., Richards, G.~T., \& Hernquist, L.\ 2007, \apj, 654, 731 
\bibitem[Hopkins et al.(2004)]{hopkins04} Hopkins, P.~F., Strauss, M.~A., Hall, P.~B., et al.\ 2004, \aj, 128, 1112 
\bibitem[Husemann et al.(2010)]{husemann10} Husemann, B., S{\'a}nchez, S.~F., Wisotzki, L., et al.\ 2010, \aap, 519, A115 
\bibitem[Ivezi{\'c} et al.(2002)]{ivezic02} Ivezi{\'c}, {\v Z}., Menou, K., Knapp, G.~R., et al.\ 2002, \aj, 124, 2364 
\bibitem[Jahnke et al.(2009)]{jahnke09} Jahnke, K., Bongiorno, A., Brusa, M., et al.\ 2009, \apjl, 706, L215 
\bibitem[Jahnke et al.(2004)]{jahnke04} Jahnke, K., S{\'a}nchez, S.~F., Wisotzki, L., et al.\ 2004, \apj, 614, 568 
\bibitem[Jahnke et al.(2007)]{jahnke07} Jahnke, K., Wisotzki, L., Courbin, F., \& Letawe, G.\ 2007, \mnras, 378, 23 
\bibitem[Jiang et al.(2009)]{jiang09} Jiang, L., Fan, X., Bian, F., et al.\ 2009, \aj, 138, 305 
\bibitem[Jiang et al.(2007)]{jiang07} Jiang, L., Fan, X., Vestergaard, M., et al.\ 2007, \aj, 134, 1150 
\bibitem[Kaiser et al.(2010)]{kaiser10} Kaiser, N., Burgett, W., Chambers, K., et al.\ 2010, \procspie, 7733, 77330E 
\bibitem[Kauffmann et al.(2003)]{kauffmann03} Kauffmann, G., Heckman, T.~M., Tremonti, C., et al.\ 2003, \mnras, 346, 1055 
\bibitem[Keel et al.(2012)]{keel12} Keel, W.~C., Chojnowski, S.~D., Bennert, V.~N., et al.\ 2012, \mnras, 420, 878 
\bibitem[Kennicutt(1998)]{kennicutt98} Kennicutt, R.~C., Jr.\ 1998, \araa, 36, 189 
\bibitem[Kewley et al.(2001)]{kewley01} Kewley, L.~J., Dopita, M.~A., Sutherland, R.~S., Heisler, C.~A., \& Trevena, J.\ 2001, \apj, 556, 121 
\bibitem[Kim et al.(2006)]{kim06} Kim, M., Ho, L.~C., \& Im, M.\ 2006, \apj, 642, 702 
\bibitem[Kirhakos et al.(1999)]{kirhakos99} Kirhakos, S., Bahcall, J.~N., Schneider, D.~P., \& Kristian, J.\ 1999, \apj, 520, 67 
\bibitem[Kormendy(1993)]{kormendy93} Kormendy, J.\ 1993, The Nearest Active Galaxies, Proceedings of the meeting on The nearest active galaxies, held in Madrid in May 1992, Madrid: Consejo Superior de Investigaciones Cientificas, 1993, Edited by J. Beckman, L. Colina and H. Netzer, p. 197
\bibitem[Kormendy \& Ho(2013)]{kormendy13} Kormendy, J., \& Ho, L.~C.\ 2013, \araa, 51, 511 
\bibitem[Kormendy \& Illingworth(1982)]{kormendy82} Kormendy, J., \& Illingworth, G.\ 1982, \apj, 256, 460 
\bibitem[Kotilainen \& Ward(1994)]{kotilainen94} Kotilainen, J.~K., \& Ward, M.~J.\ 1994, \mnras, 266, 953 
\bibitem[Krawczyk et al.(2014)]{krawczyk14} Krawczyk, C.~M., Richards, G.~T., Gallagher, S.~C., et al.\ 2014, arXiv:1412.7039 
\bibitem[Kroupa(2001)]{kroupa01} Kroupa, P.\ 2001, \mnras, 322, 231 
\bibitem[Lacy et al.(2013)]{lacy13} Lacy, M., Ridgway, S.~E., Gates, E.~L., et al.\ 2013, \apjs, 208, 24 
\bibitem[Le Borgne et al.(2003)]{leborgne03} Le Borgne, J.-F., Bruzual, G., Pell{\'o}, R., et al.\ 2003, \aap, 402, 433 
\bibitem[Liu et al.(2013a)]{liu13a} Liu, G., Zakamska, N.~L., Greene, J.~E., Nesvadba, N.~P.~H., \& Liu, X.\ 2013, \mnras, 436, 2576 
\bibitem[Liu et al.(2013b)]{liu13b} Liu, G., Zakamska, N.~L., Greene, J.~E., Nesvadba, N.~P.~H., \& Liu, X.\ 2013, \mnras, 430, 2327 
\bibitem[Liu et al.(2009)]{liu09} Liu, X., Zakamska, N.~L., Greene, J.~E., et al.\ 2009, \apj, 702, 1098 
\bibitem[Madau et al.(2014)]{madau14} Madau, P., Haardt, F., \& Dotti, M.\ 2014, \apjl, 784, LL38 
\bibitem[Magorrian et al.(1998)]{magorrian98} Magorrian, J., Tremaine, S., Richstone, D., et al.\ 1998, \aj, 115, 2285 
\bibitem[Maraston \& Str{\"o}mb{\"a}ck(2011)]{maraston11} Maraston, C., \& Str{\"o}mb{\"a}ck, G.\ 2011, \mnras, 418, 2785 
\bibitem[Marconi et al.(2004)]{marconi04} Marconi, A., Risaliti, G., Gilli, R., et al.\ 2004, \mnras, 351, 169 
\bibitem[Markwardt(2009)]{markwardt09} Markwardt, C.~B.\ 2009, Astronomical Data Analysis Software and Systems XVIII, 411, 251 
\bibitem[Matsuoka(2012)]{matsuoka12} Matsuoka, Y.\ 2012, \apj, 750, 54 
\bibitem[Matsuoka et al.(2014a)]{matsuoka14} Matsuoka, Y., Strauss, M.~A., Price, T.~N., III, \& DiDonato, M.~S.\ 2014a, \apj, 780, 162 
\bibitem[Matsuoka et al.(2014b)]{matsuoka14e} Matsuoka, Y., Strauss, M.~A., Price, T.~N., III, \& DiDonato, M.~S.\ 2014b, \apj, 789, 91 
\bibitem[Matute et al.(2012)]{matute12} Matute, I., M{\'a}rquez, I., Masegosa, J., et al.\ 2012, \aap, 542, AA20 
\bibitem[McLure \& Dunlop(2002)]{mclure02} McLure, R.~J., \& Dunlop, J.~S.\ 2002, \mnras, 331, 795 
\bibitem[McLure et al.(1999)]{mclure99} McLure, R.~J., Kukula, M.~J., Dunlop, J.~S., et al.\ 1999, \mnras, 308, 377 
\bibitem[McNamara \& Nulsen(2007)]{mcnamara07} McNamara, B.~R., \& Nulsen, P.~E.~J.\ 2007, \araa, 45, 117 
\bibitem[Merritt \& Ferrarese(2001)]{merritt01} Merritt, D., \& Ferrarese, L.\ 2001, \mnras, 320, L30 
\bibitem[Mortlock et al.(2011)]{mortlock11} Mortlock, D.~J., Warren, S.~J., Venemans, B.~P., et al.\ 2011, \nat, 474, 616 
\bibitem[Mullaney et al.(2015)]{mullaney15} Mullaney, J.~R., Alexander, D.~M., Aird, J., et al.\ 2015, arXiv:1506.05459 
\bibitem[Mullaney et al.(2012)]{mullaney12} Mullaney, J.~R., Pannella, M., Daddi, E., et al.\ 2012, \mnras, 419, 95 
\bibitem[Muzzin et al.(2013)]{muzzin13} Muzzin, A., Marchesini, D., Stefanon, M., et al.\ 2013, \apjs, 206, 8 
\bibitem[Nandra et al.(2007)]{nandra07} Nandra, K., Georgakakis, A., Willmer, C.~N.~A., et al.\ 2007, \apjl, 660, L11 
\bibitem[Nesvadba et al.(2006)]{nesvadba06} Nesvadba, N.~P.~H., Lehnert, M.~D., Eisenhauer, F., et al.\ 2006, \apj, 650, 693 
\bibitem[Netzer(2015)]{netzer15} Netzer, H.\ 2015, arXiv:1505.00811 
\bibitem[Newman et al.(2013)]{newman13} Newman, J.~A., Cooper, M.~C., Davis, M., et al.\ 2013, \apjs, 208, 5 
\bibitem[Noeske et al.(2007)]{noeske07} Noeske, K.~G., Weiner, B.~J., Faber, S.~M., et al.\ 2007, \apjl, 660, L43 
\bibitem[Oke \& Gunn(1983)]{oke83} Oke, J.~B., \& Gunn, J.~E.\ 1983, \apj, 266, 713 
\bibitem[Osterbrock \& Pogge(1985)]{osterbrock85} Osterbrock, D.~E., \& Pogge, R.~W.\ 1985, \apj, 297, 166 
\bibitem[P{\^a}ris et al.(2014)]{paris14} P{\^a}ris, I., Petitjean, P., Aubourg, {\'E}., et al.\ 2014, \aap, 563, A54 
\bibitem[Pattarakijwanich et al.(2014)]{pete15} Pattarakijwanich, P., Strauss, M.~A., Ho, S., \& Ross, N.~P.\ 2014, arXiv:1410.7394 
\bibitem[Pei(1992)]{pei92} Pei, Y.~C.\ 1992, \apj, 395, 130 
\bibitem[Peng et al.(2010)]{peng10} Peng, Y.-j., Lilly, S.~J., Kova{\v c}, K., et al.\ 2010, \apj, 721, 193 
\bibitem[Pentericci et al.(2003)]{pentericci03} Pentericci, L., Rix, H.-W., Prada, F., et al.\ 2003, \aap, 410, 75 
\bibitem[Prevot et al.(1984)]{prevot84} Prevot, M.~L., Lequeux, J., Prevot, L., Maurice, E., \& Rocca-Volmerange, B.\ 1984, \aap, 132, 389 
\bibitem[Pounds et al.(2003)]{pounds03} Pounds, K.~A., Reeves, J.~N., King, A.~R., et al.\ 2003, \mnras, 345, 705 
\bibitem[Quintero et al.(2004)]{quintero04} Quintero, A.~D., Hogg, D.~W., Blanton, M.~R., et al.\ 2004, \apj, 602, 190 
\bibitem[Reeves et al.(2009)]{reeves09} Reeves, J.~N., O'Brien, P.~T., Braito, V., et al.\ 2009, \apj, 701, 493 
\bibitem[Reyes et al.(2008)]{reyes08} Reyes, R., Zakamska, N.~L., Strauss, M.~A., et al.\ 2008, \aj, 136, 2373 
\bibitem[Richards et al.(2003)]{richards03} Richards, G.~T., Hall, P.~B., Vanden Berk, D.~E., et al.\ 2003, \aj, 126, 1131 
\bibitem[Richards et al.(2006)]{richards06} Richards, G.~T., Strauss, M.~A., Fan, X., et al.\ 2006, \aj, 131, 2766 
\bibitem[Richstone \& Schmidt(1980)]{richstone80} Richstone, D.~O., \& Schmidt, M.\ 1980, \apj, 235, 361 
\bibitem[R{\"o}nnback et al.(1996)]{ronnback96} R{\"o}nnback, J., van Groningen, E., Wanders, I., {\"O}rndahl, E.\ 1996, \mnras, 283, 282 
\bibitem[Rosario et al.(2013)]{rosario13} Rosario, D.~J., Mozena, M., Wuyts, S., et al.\ 2013, \apj, 763, 59 
\bibitem[Ross et al.(2012)]{ross12} Ross, N.~P., Myers, A.~D., Sheldon, E.~S., et al.\ 2012, \apjs, 199, 3 
\bibitem[Salim et al.(2007)]{salim07} Salim, S., Rich, R.~M., Charlot, S., et al.\ 2007, \apjs, 173, 267 
\bibitem[Salpeter(1955)]{salpeter55} Salpeter, E.~E.\ 1955, \apj, 121, 161 
\bibitem[Salvato et al.(2009)]{salvato09} Salvato, M., Hasinger, G., Ilbert, O., et al.\ 2009, \apj, 690, 1250 
\bibitem[Sanders et al.(1988)]{sanders88} Sanders, D.~B., Soifer, B.~T., Elias, J.~H., et al.\ 1988, \apj, 325, 74 
\bibitem[Santini et al.(2012)]{santini12} Santini, P., Rosario, D.~J., Shao, L., et al.\ 2012, \aap, 540, A109 
\bibitem[Schawinski et al.(2010)]{schawinski10} Schawinski, K., Urry, C.~M., Virani, S., et al.\ 2010, \apj, 711, 284 
\bibitem[Schlegel et al.(1998)]{schlegel98} Schlegel, D.~J., Finkbeiner, D.~P., \& Davis, M.\ 1998, \apj, 500, 525 
\bibitem[Schmidt \& Green(1983)]{schmidt83} Schmidt, M., \& Green, R.~F.\ 1983, \apj, 269, 352 
\bibitem[Schneider et al.(2010)]{schneider10} Schneider, D.~P., Richards, G.~T., Hall, P.~B., et al.\ 2010, \aj, 139, 2360 
\bibitem[Schulze \& Wisotzki(2011)]{schulze11} Schulze, A., \& Wisotzki, L.\ 2011, \aap, 535, A87 
\bibitem[S\'{e}rsic(1968)]{sersic68} Sersic, J.~L.\ 1968, Cordoba, Argentina: Observatorio Astronomico, 1968
\bibitem[Shen(2013)]{shen13} Shen, Y.\ 2013, Bulletin of the Astronomical Society of India, 41, 61 
\bibitem[Shen et al.(2015a)]{shen14} Shen, Y., Brandt, W.~N., Dawson, K.~S., et al.\ 2015, \apjs, 216, 4 
\bibitem[Shen et al.(2015b)]{shen15} Shen, Y., Greene, J.~E., Ho, L.~C., et al.\ 2015, arXiv:1502.01034 
\bibitem[Shen \& Liu(2012)]{shen12} Shen, Y., \& Liu, X.\ 2012, \apj, 753, 125 
\bibitem[Shen et al.(2011)]{shen11} Shen, Y., Richards, G.~T., Strauss, M.~A., et al.\ 2011, \apjs, 194, 45 
\bibitem[Shi et al.(2009)]{shi09} Shi, Y., Rieke, G.~H., Ogle, P., Jiang, L., \& Diamond-Stanic, A.~M.\ 2009, \apj, 703, 1107 
\bibitem[Silverman et al.(2009)]{silverman09} Silverman, J.~D., Lamareille, F., Maier, C., et al.\ 2009, \apj, 696, 396 
\bibitem[Silverman et al.(2008)]{silverman08} Silverman, J.~D., Mainieri, V., Lehmer, B.~D., et al.\ 2008, \apj, 675, 1025 
\bibitem[Smee et al.(2013)]{smee13} Smee, S.~A., Gunn, J.~E., Uomoto, A., et al.\ 2013, \aj, 146, 32 
\bibitem[Soltan(1982)]{soltan82} Soltan, A.\ 1982, \mnras, 200, 115 
\bibitem[Somerville et al.(2008)]{somerville08} Somerville, R.~S., Hopkins, P.~F., Cox, T.~J., Robertson, B.~E., \& Hernquist, L.\ 2008, \mnras, 391, 481 
\bibitem[Springel et al.(2005)]{springel05} Springel, V., Di Matteo, T., \& Hernquist, L.\ 2005, \apjl, 620, L79 
\bibitem[Steinhardt et al.(2014)]{steinhardt14} Steinhardt, C.~L., Speagle, J.~S., Capak, P., et al.\ 2014, \apjl, 791, L25 
\bibitem[Sun et al.(2015)]{sun15} Sun, M., Trump, J.~R., Brandt, W.~N., et al.\ 2015, \apj, 802, 14 
\bibitem[Takada et al.(2014)]{takada14} Takada, M., Ellis, R.~S., Chiba, M., et al.\ 2014, \pasj, 66, R1 
\bibitem[Tombesi et al.(2010)]{tombesi10} Tombesi, F., Cappi, M., Reeves, J.~N., et al.\ 2010, \aap, 521, A57 
\bibitem[Trump et al.(2006)]{trump06} Trump, J.~R., Hall, P.~B., Reichard, T.~A., et al.\ 2006, \apjs, 165, 1 
\bibitem[Trump et al.(2013)]{trump13} Trump, J.~R., Hsu, A.~D., Fang, J.~J., et al.\ 2013, \apj, 763, 133 
\bibitem[Trump et al.(2015)]{trump15} Trump, J.~R., Sun, M., Zeimann, G.~R., et al.\ 2015, arXiv:1501.02801 
\bibitem[Urrutia et al.(2008)]{urrutia08} Urrutia, T., Lacy, M., \& Becker, R.~H.\ 2008, \apj, 674, 80 
\bibitem[Vanden Berk et al.(2001)]{vandenberk01} Vanden Berk, D.~E., Richards, G.~T., Bauer, A., et al.\ 2001, \aj, 122, 549 
\bibitem[Vanden Berk et al.(2006)]{vandenberk06} Vanden Berk, D.~E., Shen, J., Yip, C.-W., et al.\ 2006, \aj, 131, 84 
\bibitem[V{\'e}ron-Cetty et al.(2004)]{veron-cetty04} V{\'e}ron-Cetty, M.-P., Joly, M., \& V{\'e}ron, P.\ 2004, \aap, 417, 515 
\bibitem[Vestergaard \& Peterson(2006)]{vestergaard06} Vestergaard, M., \& Peterson, B.~M.\ 2006, \apj, 641, 689 
\bibitem[Wang et al.(2010)]{wang10} Wang, R., Carilli, C.~L., Neri, R., et al.\ 2010, \apj, 714, 699 
\bibitem[Willott et al.(2010)]{willott10} Willott, C.~J., Delorme, P., Reyl{\'e}, C., et al.\ 2010, \aj, 139, 906 
\bibitem[Wu et al.(2015)]{wu15} Wu, X.-B., Wang, F., Fan, X., et al.\ 2015, \nat, 518, 512 
\bibitem[Xue et al.(2010)]{xue10} Xue, Y.~Q., Brandt, W.~N., Luo, B., et al.\ 2010, \apj, 720, 368 
\bibitem[Yesuf et al.(2014)]{yesuf14} Yesuf, H.~M., Faber, S.~M., Trump, J.~R., et al.\ 2014, \apj, 792, 84 
\bibitem[York et al.(2000)]{york00} York, D.~G., Adelman, J., Anderson, J.~E., Jr., et al.\ 2000, \aj, 120, 1579 
\bibitem[Yu \& Tremaine(2002)]{yu02} Yu, Q., \& Tremaine, S.\ 2002, \mnras, 335, 965 
\bibitem[Zakamska et al.(2008)]{zakamska08} Zakamska, N.~L., G{\'o}mez, L., Strauss, M.~A., \& Krolik, J.~H.\ 2008, \aj, 136, 1607 
\bibitem[Zakamska et al.(2006)]{zakamska06} Zakamska, N.~L., Strauss, M.~A., Krolik, J.~H., et al.\ 2006, \aj, 132, 1496 
\bibitem[Zakamska et al.(2003)]{zakamska03} Zakamska, N.~L., Strauss, M.~A., Krolik, J.~H., et al.\ 2003, \aj, 126, 2125 
\end{thebibliography}
\end{document}